\documentclass[a4paper,11pt]{article}%
\usepackage{amsmath}
\usepackage{epsf}
\usepackage{graphicx}
\usepackage{cite}%
\usepackage{amsfonts}%
\usepackage{amssymb}
\begin{document}
\thispagestyle{empty}  \setcounter{page}{0}  \begin{flushright}%

June 2006 \\
TTP06-19\\
\end{flushright}%

\vskip                          3.5 true cm

\begin{center}
{\huge $K_{L}\rightarrow\pi^{0}e^{+}e^{-}$ and $K_{L}\rightarrow\pi^{0}\mu
^{+}\mu^{-}$ :\smallskip\smallskip}

{\huge A binary star on the stage of flavor physics}\\[28pt]\textsc{Federico
Mescia},$^{1}$ \textsc{Christopher Smith},$^{2}$ \textsc{Stephanie Trine}%
${}^{3}$ \\[25pt]$^{1}~$\textsl{INFN, Laboratori Nazionali di Frascati,
I-00044 Frascati, Italy}\\[5pt]$^{2}~$\textsl{Institut f\"{u}r Theoretische
Physik, Universit\"{a}t Bern, CH-3012 Bern, Switzerland }\\[5pt]$^{3}%
~$\textsl{Institut f\"{u}r Theoretische Teilchenphysik,}
\textsl{Universit\"{a}t Karlsruhe, D-76128 Karlsruhe, Germany }\\[45pt]

\textbf{Abstract}
\end{center}

\noindent A systematic analysis of New Physics impacts on the rare decays
$K_{L}\rightarrow\pi^{0}\ell^{+}\ell^{-}$ is performed. Thanks to their
different sensitivities to flavor-changing local effective interactions, these
two modes could provide valuable information on the nature of the possible New
Physics at play. In particular, a combined measurement of both modes could
disentangle scalar/pseudoscalar from vector or axial-vector contributions. For
the latter, model-independent bounds are derived. Finally, the $K_{L}%
\rightarrow\pi^{0}\mu^{+}\mu^{-}$ forward-backward CP-asymmetry is considered,
and shown to give interesting complementary information.\newpage

\section{Introduction}

Rare $K$ decays, directly sensitive to short-distance FCNC processes, offer an
invaluable window into the physics at play at high-energy scales. Besides the
two $K\rightarrow\pi\nu\bar{\nu}$ golden modes, the decays $K_{L}%
\rightarrow\pi^{0}e^{+}e^{-}$ and $K_{L}\rightarrow\pi^{0}\mu^{+}\mu^{-}$ also
exhibit good sensitivities, thanks to the theoretical control achieved over
their long-distance components\cite{DEIP,BDI03,IsidoriSmithUnter}.
Importantly, these modes are sensitive to different combinations of
short-distance FCNC currents, and thus allow in principle to discriminate
among possible New Physics scenarios.

In this respect, the pair of $K_{L}\rightarrow\pi^{0}\ell^{+}\ell^{-}$ decays
is unique since, though their dynamics is similar, the very different lepton
masses allow to probe helicity-suppressed effects in a particularly clean way.
Only the $K_{L}\rightarrow\ell^{+}\ell^{-}$ modes share this characteristic,
but the dominance of the long-distance two-photon contribution unfortunately
prevents from acceding to the short-distance physics with a good degree of
precision\cite{IsidoriU03}. On the contrary, the corresponding two-photon
contribution to $K_{L}\rightarrow\pi^{0}\ell^{+}\ell^{-}$ is under control. It
represents only $30\%$ of the total rate for the muonic mode, and is
negligible for the electronic one\cite{BDI03,IsidoriSmithUnter}.

The main purpose of the paper is to illustrate how this fact can be used to
constrain or identify the nature of possible New Physics effects. More
precisely, our goals are:

\begin{enumerate}
\item To analyze the impacts arising from all possible $\Delta S=1$
four-fermion operators in a model-independent way, i.e. operators of the form
$\left(  \bar{s}\Gamma d\right)  \left(  \bar{\ell}\Gamma\ell\right)  $,
$\Gamma=1,\gamma_{5},\gamma^{\mu},\gamma^{\mu}\gamma_{5},\sigma^{\mu\nu}$, and
to show how combined measurements of $K_{L}\rightarrow\pi^{0}\ell^{+}\ell^{-}$
can disentangle them. An important distinction is made between
helicity-suppressed operators, like the $\Gamma=1,\gamma_{5}$ ones arising for
example in the MSSM at large $\tan\beta$\cite{TanBeta,Foster05}, and
helicity-allowed operators like in SUSY without R parity\cite{Rparity} or from
leptoquark interactions\cite{DavidsonBC94}. Also, the electromagnetic tensor
operator $\bar{s}\sigma_{\mu\nu}dF^{\mu\nu}$ will be briefly
considered\cite{BurasCIRS99,DAmbrosioGao}. Finally, constraints from
$K_{L}\rightarrow\ell^{+}\ell^{-}$ (for scalar/pseudoscalar operators) and
$K\rightarrow\pi\nu\bar{\nu}$ (for helicity-allowed tensor/pseudotensor
interactions) will be analyzed.

\item To improve the control over the long-distance two-photon contribution.
Indeed, this is needed to estimate interference effects with New Physics
short-distance contributions. In addition, once achieved, the $K_{L}%
\rightarrow\pi^{0}\mu^{+}\mu^{-}$ forward-backward (or lepton energy)
CP-asymmetry\cite{2pp,AFB} will be computed reliably for the first time, both
in the Standard Model and beyond, and will prove to be an interesting
complementary source of information on the New Physics at play.
\end{enumerate}

The paper is organized as follows. In Section 2, we present the ingredients
needed to deal with the long-distance dominated contributions, and, in the
spirit of Ref.\cite{IsidoriSmithUnter}, analyze the possible signals of New
Physics in the vector or axial-vector operators. Then, in Section 3, we
analyze all other operators, both in the helicity-suppressed and
helicity-allowed cases. The corresponding analysis of $K_{L,S}\rightarrow
\ell^{+}\ell^{-}$ is in appendix. Finally, our results are summarized in the Conclusion.

\section{$K_{L}\rightarrow\pi^{0}\ell^{+}\ell^{-}$ with standard
short-distance operators}

The $K_{L}\rightarrow\pi^{0}\ell^{+}\ell^{-}$ decays receive essentially three
types of contributions, depicted in Fig.\ref{Fig1}.

A first class of effects, purely sensitive to short-distance physics, results
from heavy particle FCNC loops (the $W$ and $Z$ bosons and the $t$ and $c$
quarks in the SM, see Fig.\ref{Fig1}a), and can be parametrized by a set of
local effective operators. In the SM, the leading relevant effective
Hamiltonian induced by these effects reads \cite{BuchallaBL96}:%
\begin{equation}
\mathcal{H}_{eff}^{\mathrm{V,A}}=-\frac{G_{F}\alpha}{\sqrt{2}}\lambda
_{t}\left[  y_{7V}\,\left(  \bar{s}\gamma_{\mu}d\right)  \left(  \bar{\ell
}\gamma^{\mu}\ell\right)  +y_{7A}\left(  \bar{s}\gamma_{\mu}d\right)  \left(
\bar{\ell}\gamma^{\mu}\gamma_{5}\ell\right)  \right]  +h.c.\;, \label{Eq2}%
\end{equation}
with $\alpha\equiv\alpha\left(  M_{Z}\right)  $ and $\lambda_{q}=V_{qs}^{\ast
}V_{qd}$. Of course, in the presence of New Physics, other types of effective
interactions could be produced. This will be the subject of Section 3. For
now, we will assume that New Physics affects only the values of the
coefficients $y_{7A,7V}$, and leave them as free
parameters\cite{IsidoriSmithUnter}.

Beyond SM scenarios leading to such modifications of the vector and
axial-vector couplings are numerous. Examples are the MSSM for moderate values
of $\tan\beta$ (see e.g. Ref.\cite{MSSM}, and references therein), for large
$\tan\beta$ (from charged Higgs penguins, see e.g. Ref.\cite{isidoriP06}), or
the Enhanced Electroweak Penguins of Refs.\cite{BurasS99,BFRS04}. Of course,
in specific models, New Physics also affects operators with different flavor
quantum numbers. We have opted here for a decoupled, model-independent
analysis, considering thus only the operators relevant for $K_{L}%
\rightarrow\pi^{0}\ell^{+}\ell^{-}$.

Also, the four-quark operators $Q_{1,...,6}$ have not been explicitly included
in Eq.(\ref{Eq2}). This is because their impact on the direct CP-violating
(DCPV) contribution to $K_{L}\rightarrow\pi^{0}\ell^{+}\ell^{-}$, associated
to the local effective Hamiltonian $\mathcal{H}_{eff}^{\mathrm{V,A}}$ in
standard terminology, can be safely neglected in the SM
\cite{BuchallaBL96,BDI03}. New Physics cannot change this picture as new
sources of CP-violation from $Q_{1,...,6}$ are bounded from purely hadronic
$K$ decay observables.

A second class of contributions, dominated by long-distance dynamics, is
driven by the coupling of leptons to photons, via $K^{0}-\bar{K}^{0}$ mixing
(Fig.\ref{Fig1}b) or via a two photon loop (Fig.\ref{Fig1}c). Let us now
analyze these in more detail.

\subsection{Long-distance dominated contributions}

The remarkable point with the contributions depicted in Fig.1b and 1c is that
they can be entirely determined from experimental data. Their estimation is
thus not affected by possible New Physics effects. As they remain as an
unavoidable background to the interesting short-distance contributions, we
briefly recall (and partly improve, in the case of two-photon amplitudes) the
way they are dealt with.%

\begin{figure}
[t]
\begin{center}
\includegraphics[
height=2.2632in,
width=4.6103in
]%
{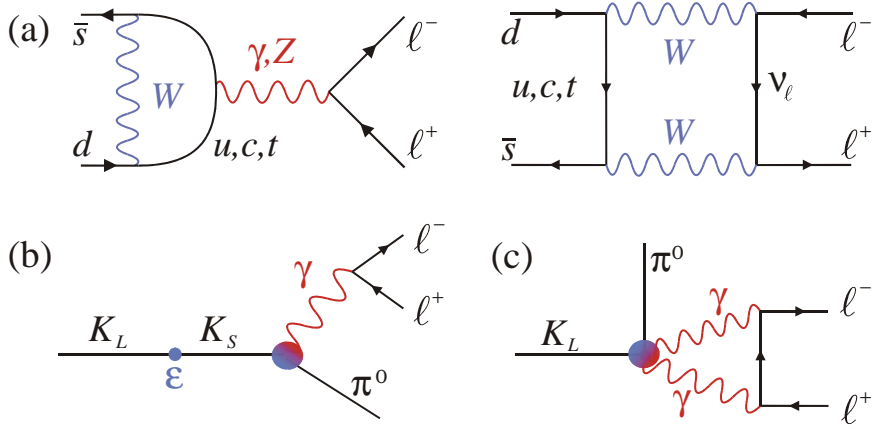}%
\caption{(a) Short-distance penguin and box diagrams for the initial
conditions of $y_{7A,7V}$ in the SM. (b) $K^{0}-\bar{K}^{0}$ mixing-induced
contribution, with a long-distance dominated, CP-conserving effective
$K_{S}\rightarrow\pi^{0}\gamma^{\ast}$ vertex. (c) Long-distance two-photon
induced CP-conserving contributions.}%
\label{Fig1}%
\end{center}
\end{figure}

\paragraph{\textit{The indirect CP-violating contribution}}

(ICPV, Fig.\ref{Fig1}b) originates from $K^{0}-\bar{K}^{0}$ mixing. The
subsequent CP-conserving $K_{S}\rightarrow\pi^{0}\ell^{+}\ell^{-}$ decay is
dominated by the long-distance process $K_{S}\rightarrow\pi^{0}\left(
\gamma^{\ast}\rightarrow\ell^{+}\ell^{-}\right)  $, producing the lepton pair
in a $1^{--}$ state. The corresponding amplitude reads:%
\begin{equation}
\mathcal{M}_{\mathrm{ICPV}}=\varepsilon\,\mathcal{M}\left(  K_{1}%
\rightarrow\pi^{0}\ell^{+}\ell^{-}\right)  =-\varepsilon\frac{G_{F}\alpha
_{em}}{4\pi}W_{S}\left(  z\right)  \left(  P+K\right)  ^{\mu}\left\{
\bar{u}_{p}\gamma_{\mu}v_{p^{\prime}}\right\}  \;, \label{Eq7}%
\end{equation}
where $P,K,p$ and $p^{\prime}$ denote the momenta of the $K,\pi,\ell^{-}$ and
$\ell^{+}$ states, respectively, $z=T^{2}/M_{K^{0}}^{2}$ with $T=p+p^{\prime}$
and $\alpha_{em}\approx1/137$. The $W_{S}$ function has been analyzed in
detail in Chiral Perturbation Theory (ChPT) in Ref.\cite{DEIP}, and can be
parametrized as follows:
\begin{equation}
W_{S}\left(  z\right)  =a_{S}+b_{S}z+W_{S}^{\pi\pi}\left(  z\right)
+W_{S}^{KK}\left(  z\right)  \;. \label{Eq8}%
\end{equation}
The pion and kaon loops ($W_{S}^{\pi\pi},W_{S}^{KK}$) were found small and, to
a good approximation, a single (real) counterterm dominates: $W_{S}\left(
z\right)  \approx a_{S}$. This counterterm can then be extracted from the
experimental $K_{S}\rightarrow\pi^{0}e^{+}e^{-}$ and $K_{S}\rightarrow\pi
^{0}\mu^{+}\mu^{-}$ branching fractions: $\left|  a_{S}\right|  =1.20\pm0.20$
\cite{KSpill}. Eq.(\ref{Eq7}) is thus indeed entirely determined in terms of
measured quantities ($|a_{S}|$, $\varepsilon$).

\paragraph{\textit{The two-photon CP-conserving contribution}}

(Fig.\ref{Fig1}c), $K_{L}\rightarrow\pi^{0}\left(  \gamma^{\ast}\gamma^{\ast
}\rightarrow\ell^{+}\ell^{-}\right)  $, produces the lepton pair in either a
phase-space suppressed tensor state $2^{++}$ or a helicity-suppressed scalar
state $0^{++}$. The former is found negligible from experimental constraints
on $K_{L}\rightarrow\pi^{0}\left(  \gamma\gamma\right)  _{2^{++}}$%
\cite{BDI03}, while for the latter the amplitude reads:%
\begin{equation}
\mathcal{M}_{\mathrm{\gamma\gamma}}=\frac{G_{8}\alpha_{em}^{2}}{2\pi^{2}}%
M_{K}\frac{r_{\ell}}{z}\mathcal{F}_{\ell}\left(  z\right)  \left\{
\bar{u}_{p}v_{p^{\prime}}\right\}  \;, \label{Eq8b}%
\end{equation}
with $r_{\ell}=m_{\ell}/M_{K}$ and $\left|  G_{8}\right|  =9.1\cdot10^{-12}$
MeV$^{-2}$. It is dominated by the two-loop process $K_{L}\rightarrow\pi
^{0}\left(  P^{+}P^{-}\rightarrow\gamma^{\ast}\gamma^{\ast}\rightarrow\ell
^{+}\ell^{-}\right)  _{0^{++}}$ with $P=\pi,K$, computed in ChPT, and closely
related to $K_{S}\rightarrow\pi^{+}\pi^{-}\rightarrow\gamma^{\ast}\gamma
^{\ast}\rightarrow\ell^{+}\ell^{-}$. With the parametrization $\mathcal{M}%
\left(  K_{L}\rightarrow\pi^{0}P^{+}P^{-}\right)  \sim a_{1}^{P}\left(
z\right)  $ for the momentum distribution entering the subprocess $P^{+}%
P^{-}\rightarrow\gamma^{\ast}\gamma^{\ast}\rightarrow\ell^{+}\ell^{-}$, the
two-loop form-factor $\mathcal{F}_{\ell}\left(  z\right)  $ can be expressed
as ($r_{\pi}=M_{\pi}/M_{K}$)%
\begin{equation}
\mathcal{F}_{\ell}\left(  z\right)  =a_{1}^{\pi}\left(  z\right)
\mathcal{I}\left(  \frac{r_{\ell}^{2}}{z},\frac{r_{\pi}^{2}}{z}\right)
-a_{1}^{K}\left(  z\right)  \mathcal{I}\left(  \frac{r_{\ell}^{2}}{z}%
,\frac{1}{z}\right)  \;, \label{Eq9a}%
\end{equation}
with $\mathcal{I}\left(  a,b\right)  $ given in
Refs.\cite{EckerPich91,IsidoriSmithUnter} (for practical purposes, a numerical
representation is given in Appendix B).

A reliable estimation of $\Gamma\left(  K_{L}\rightarrow\pi^{0}\ell^{+}%
\ell^{-}\right)  _{\mathrm{\gamma\gamma}}$ can then be obtained from the
measured $K_{L}\rightarrow\pi^{0}\gamma\gamma$ rate\cite{ggrate} thanks to the
stability of the ratio%
\begin{equation}
R_{\mathrm{\gamma\gamma}}=\frac{\Gamma\left(  K_{L}\rightarrow\pi^{0}\ell
^{+}\ell^{-}\right)  _{\mathrm{\gamma\gamma}}}{\Gamma\left(  K_{L}%
\rightarrow\pi^{0}\gamma\gamma\right)  }\;, \label{Eq9}%
\end{equation}
with respect to changes in $a_{1}^{P}\left(  z\right)  $
\cite{IsidoriSmithUnter}. Note that some $\mathcal{O}\left(  p^{6}\right)  $
ChPT effects are thus included in the estimated $\Gamma\left(  K_{L}%
\rightarrow\pi^{0}\ell^{+}\ell^{-}\right)  _{\mathrm{\gamma\gamma}}$, most
notably those responsible for the large observed $K_{L}\rightarrow\pi
^{0}\gamma\gamma$ rate compared to the $\mathcal{O}\left(  p^{4}\right)  $
ChPT prediction.

However, theoretical control over $R_{\mathrm{\gamma\gamma}}$ only is not
sufficient to deal with New Physics interactions that produce the lepton pair
in a $0^{++}$ state (generating interference effects) or compute
forward-backward asymmetries. For this, we need to control $\mathcal{M}%
_{\mathrm{\gamma\gamma}}$ too. To fill this gap, the key is to use the
similarity of behavior of the $K_{L}\rightarrow\pi^{0}\gamma\gamma$ and
$K_{L}\rightarrow\pi^{0}\ell^{+}\ell^{-}$ spectra at the origin of the
stability of $R_{\mathrm{\gamma\gamma}}$. First consider the fact that the
$K_{L}\rightarrow\pi^{0}\gamma\gamma$ normalized spectrum is rather
well-described with the parametrizations
\begin{subequations}
\label{Eq10}%
\begin{align}
\mathcal{O}\left(  p^{4}\right)  \;\text{ChPT}  &  :a_{1}^{\pi}\left(
z\right)  _{ChPT}=z-r_{\pi}^{2},\\
K_{L}\overset{}{\rightarrow}\pi^{0}\pi^{+}\pi^{-}\text{ Dalitz}  &
:a_{1}^{\pi}\left(  z\right)  _{Dalitz}=-0.46+2.44z-0.95z^{2}\,,
\end{align}
and $a_{1}^{K}\left(  z\right)  =z-1-r_{\pi}^{2}$ in both cases. For
$a_{1}^{\pi}\left(  z\right)  _{Dalitz}$, only the $z$-dependent part of the
$K_{L}\rightarrow\pi^{0}\pi^{+}\pi^{-}$ effective vertex is kept. Now, to
account for the rescaling of $\mathcal{B}\left(  K_{L}\rightarrow\pi^{0}%
\ell^{+}\ell^{-}\right)  _{\mathrm{\gamma\gamma}}$ by $\mathcal{B}\left(
K_{L}\rightarrow\pi^{0}\gamma\gamma\right)  ^{\exp}$, we rescale $a_{1}%
^{P}\left(  z\right)  _{ChPT}\rightarrow1.45a_{1}^{P}\left(  z\right)
_{ChPT}$ and $a_{1}^{P}\left(  z\right)  _{Dalitz}\rightarrow1.36a_{1}%
^{P}\left(  z\right)  _{Dalitz}$, such that the rate computed from
Eq.(\ref{Eq8b}) is kept frozen. Clearly, the theoretical control on the
resulting $K_{L}\rightarrow\pi^{0}\ell^{+}\ell^{-}$ differential rate is not
as good as on the total rate, but is nevertheless satisfactory. In practice,
the error inherent to the procedure can be probed by comparing the predictions
obtained using either $a_{1}^{\pi}\left(  z\right)  _{ChPT}$ or $a_{1}^{\pi
}\left(  z\right)  _{Dalitz}$.

Finally, as far as the sign of $\mathcal{M}_{\mathrm{\gamma\gamma}}$ is
concerned, we checked that Eq.(\ref{Eq8b}) is consistent with the conventions
used in the rest of the paper for hadronic matrix elements. Furthermore, under
the reasonable assumption that the sign of $G_{8}$ as fixed by the
factorization approximation is not changed by the non-perturbative evolution
down to $M_{K}$ (see for example Ref.\cite{GerardST}), our conventions
correspond to $G_{8}<0$.

\subsection{Vector and axial-vector short-distance contributions}

Let us now consider the DCPV piece induced by the vector and axial-vector
operators of Eq.(\ref{Eq2}):%
\end{subequations}
\begin{equation}
\mathcal{M}_{\mathrm{V,A}}=\langle\pi^{0}\ell^{+}\ell^{-}|-\mathcal{H}%
_{eff}^{\mathrm{V,A}}|K_{L}\rangle=iG_{F}\alpha\langle\pi^{0}|\bar{s}%
\gamma_{\mu}d|K^{0}\rangle\left\{  \bar{u}_{p}\gamma^{\mu}\left(
\operatorname{Im}\left(  \lambda_{t}y_{7V}\right)  +\operatorname{Im}\left(
\lambda_{t}y_{7A}\right)  \gamma_{5}\right)  v_{p^{\prime}}\right\}
\label{Eq3b}%
\end{equation}
(the sizeable $c$-quark contribution, known at NLO\cite{BuchallaBL96}, is
understood in $y_{7V}$), with the matrix element%
\begin{gather}
\langle\pi^{0}\left(  K\right)  |\bar{s}\gamma^{\mu}d|K^{0}\left(  P\right)
\rangle=\frac{1}{\sqrt{2}}\left(  \left(  P+K\right)  ^{\mu}f_{+}^{K^{0}%
\pi^{0}}\left(  z\right)  +\left(  P-K\right)  ^{\mu}f_{-}^{K^{0}\pi^{0}%
}\left(  z\right)  \right)  \;,\label{Eq4}\\
f_{-}^{K^{0}\pi^{0}}\left(  z\right)  =\frac{1-r_{\pi}^{2}}{z}\left(
f_{0}^{K^{0}\pi^{0}}\left(  z\right)  -f_{+}^{K^{0}\pi^{0}}\left(  z\right)
\right)  \;.\nonumber
\end{gather}
The slopes are extracted from $K_{\ell3}$ decays \cite{KL3,PDG}, neglecting
isospin breaking:%
\begin{equation}
f_{0,+}\left(  z\right)  \equiv f_{0,+}^{K^{0}\pi^{0}}\left(  z\right)
=\frac{f_{+}\left(  0\right)  }{1-\lambda_{0,+}z},\;\;\lambda_{0}=0.18\text{,
}\lambda_{+}=0.32\;, \label{Eq5}%
\end{equation}
in the pole parametrization. Accounting for isospin breaking in $\pi^{0}-\eta$
mixing for the value at zero momentum transfer\cite{MarcianoParsa}, one gets:%
\begin{equation}
f_{+}\left(  0\right)  =\left(  1.0231\right)  ^{-1}f_{+}^{K^{0}\pi^{+}%
}\left(  0\right)  \approx0.939\;, \label{Eq6}%
\end{equation}
with the Leutwyler-Ross prediction $f_{+}^{K^{0}\pi^{+}}\left(  0\right)
=0.961\left(  8\right)  $\cite{LeutwylerRoos}, confirmed by lattice
studies\cite{LatticeF0}. This quite precise value, together with the knowledge
of the form-factor slopes, renders the theoretical prediction of the vector
and axial-vector contributions to $K_{L}\rightarrow\pi^{0}\ell^{+}\ell^{-}$
remarkably clean.

Inserting Eq.(\ref{Eq4}) in $\mathcal{M}_{\mathrm{V,A}}$, the vector current
is seen to produce the lepton pair in a vector state $1^{--}$, while the
axial-vector part produces it in an axial-vector $1^{++}$ or pseudoscalar
$0^{-+}$ state. This latter component is helicity suppressed, and enters thus
only the muon mode. Besides, the vector current and ICPV amplitudes produce
the same final state, they thus interfere in the rate. Recent theoretical
analyses point towards a constructive interference, i.e., $a_{S}<0$
\cite{BDI03,FGD04}.%

\begin{figure}
[t]
\begin{center}
\includegraphics[
height=2.808in,
width=3.8009in
]%
{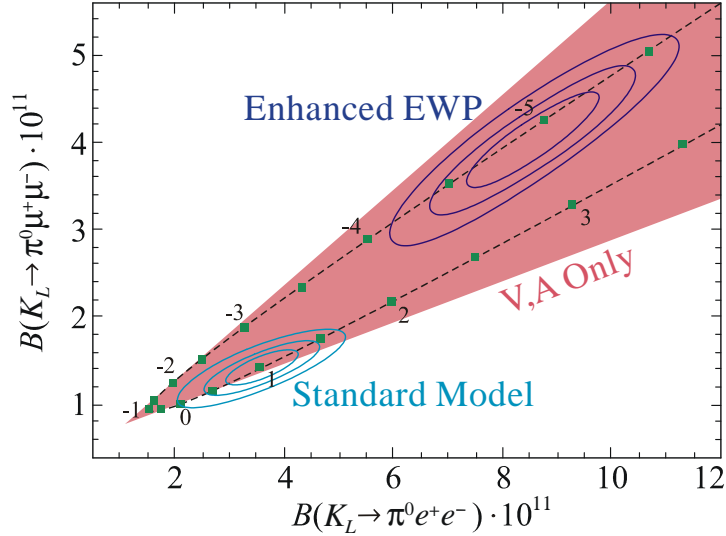}%
\caption{Behavior of $\mathcal{B}\left(  K_{L}\rightarrow\pi^{0}\mu^{+}\mu
^{-}\right)  $ against $\mathcal{B}\left(  K_{L}\rightarrow\pi^{0}e^{+}%
e^{-}\right)  $, in units of $10^{-11}$, as $y_{7A,7V}$ are rescaled by a
common factor (dots), or allowed to take arbitrary values (red sector). The
ellipses denote $25,50,75\%$ confidence regions in the SM, assuming
constructive DCPV -- ICPV interference (destructive interference is around the
$-1$ dot), or for Ref.\cite{BFRS04}, in which $y_{7A}\approx-3.2$ and
$y_{7V}\approx0.9$.}%
\label{Fig2}%
\end{center}
\end{figure}

\paragraph{\textit{Total rates}:}

Altogether, the branching ratios are predicted to be%
\begin{gather}
\mathcal{B}_{\mathrm{V,A}}^{\ell^{+}\ell^{-}}=\left(  C_{dir}^{\ell}\pm
C_{int}^{\ell}\left|  a_{S}\right|  +C_{mix}^{\ell}\left|  a_{S}\right|
^{2}+C_{\gamma\gamma}^{\ell}\right)  \cdot10^{-12}\;,\smallskip\label{Eq11}\\%
\begin{tabular}
[c]{ll}%
$C_{dir}^{e}=\left(  4.62\pm0.24\right)  \;\left(  w_{7V}^{2}+w_{7A}%
^{2}\right)  ,$ & $C_{dir}^{\mu}=\left(  1.09\pm0.05\right)  \left(
w_{7V}^{2}+2.32w_{7A}^{2}\right)  ,$\\
$C_{int}^{e}=\left(  11.3\pm0.3\right)  \;w_{7V},$ & $C_{int}^{\mu}=\left(
2.63\pm0.06\right)  \;w_{7V},$\\
$C_{mix}^{e}=14.5\pm0.5,$ & $C_{mix}^{\mu}=3.36\pm0.20,$\\
$C_{\gamma\gamma}^{e}\approx0,$ & $C_{\gamma\gamma}^{\mu}=5.2\pm1.6,$%
\end{tabular}
\nonumber
\end{gather}
with $w_{7A,7V}=\operatorname{Im}\left(  \lambda_{t}y_{7A,7V}\right)
/\operatorname{Im}\lambda_{t}$. In the Standard Model, the coefficients
$y_{7A,7V}$ are real \cite{BuchallaBL96}:%
\begin{equation}
y_{7A}\left(  M_{W}\right)  =-0.68\pm0.03,\;\;y_{7V}\left(  \mu\approx
1\;\text{GeV}\right)  =0.73\pm0.04\;, \label{Eq3}%
\end{equation}
and, with $\operatorname{Im}\lambda_{t}=\left(  1.407\pm0.098\right)
\cdot10^{-4}$\cite{CKMfitter}, one gets%
\begin{equation}
\mathcal{B}_{\mathrm{SM}}^{e^{+}e^{-}}=3.54_{-0.85}^{+0.98}\;\left(
1.56_{-0.49}^{+0.62}\right)  \cdot10^{-11},\;\;\mathcal{B}_{\mathrm{SM}}%
^{\mu^{+}\mu^{-}}=1.41_{-0.26}^{+0.28}\;\left(  0.95_{-0.21}^{+0.22}\right)
\cdot10^{-11}\;, \label{Eq11b}%
\end{equation}
for constructive (destructive) interference. The present experimental bounds
are one order of magnitude above these predictions:%
\begin{equation}
\mathcal{B}_{\mathrm{\exp}}^{e^{+}e^{-}}<28\cdot10^{-11}\text{\cite{KTeVelec}%
},\;\;\mathcal{B}_{\mathrm{\exp}}^{\mu^{+}\mu^{-}}<38\cdot10^{-11}%
\text{\cite{KTeVmuon}}\;. \label{Eq11c}%
\end{equation}
The differential rate for the electronic mode is trivial (no CPC piece), while
for the muonic mode it can be found in Ref.\cite{IsidoriSmithUnter}.

The Standard Model confidence region on the $\mathcal{B}^{e^{+}e^{-}}%
$--$\mathcal{B}^{\mu^{+}\mu^{-}}$ plane is shown in Fig.\ref{Fig2}. As
discussed in Ref.\cite{IsidoriSmithUnter}, this plane is particularly
well-suited to search for New Physics signal, and identify its specific
nature. Indeed, the fact that helicity suppression is rather inefficient for
the muonic mode introduces a genuine difference of sensitivity to the
short-distance V,A currents, i.e. to $y_{7A,7V}$, for the two $K_{L}%
\rightarrow\pi^{0}\ell^{+}\ell^{-}$ modes. These two types of contributions
can thus be disentangled by measuring both rates. Suffices to note that the
coefficients approximately obey $C_{i}^{\mu}/C_{i}^{e}\approx0.23$,
$i=dir,int,mix$, which is simply the phase-space suppression, except for the
enhanced $y_{7A}$ contribution to $C_{dir}^{\mu}$, which comes from the
production of helicity-suppressed pseudoscalar $0^{-+}$ states. Without this,
no matter the New Physics contributions to $y_{7A,7V}$, the rates would always
fall on a trivial straight line. Thanks to this effect, on the contrary, for
arbitrary values of $y_{7A,7V}$, the two modes can lie anywhere inside the red
sector in Fig.\ref{Fig2} for the $C_{i}^{\ell}$ at their central values.
Accounting for theoretical errors at $1\sigma$, this area is mathematically
expressed as%
\begin{equation}
0.1\cdot10^{-11}+0.24\mathcal{B}_{\mathrm{V,A}}^{e^{+}e^{-}}<\mathcal{B}%
_{\mathrm{V,A}}^{\mu^{+}\mu^{-}}<0.6\cdot10^{-11}+0.58\mathcal{B}%
_{\mathrm{V,A}}^{e^{+}e^{-}}\;. \label{Eq11d}%
\end{equation}
For definiteness, we have also indicated the curve corresponding to a common
rescaling of $y_{7A}$ and $y_{7V}$ from New Physics, and drawn the confidence
region for the enhanced electroweak penguins of Ref.\cite{BFRS04}, to
illustrate the opposite situation in which $y_{7A}=-3.2$ is strongly enhanced,
while $y_{7V}=0.9$ stays roughly the same as in the SM. Finally, note that the
extent of the confidence regions essentially reflects the uncertainty on
$a_{S}$ (whose effect is included in the bounds Eq.(\ref{Eq11d})), and could
be reduced by more precise measurements of the $K_{S}\rightarrow\pi^{0}%
\ell^{+}\ell^{-}$ branching fractions.%

\begin{figure}
[t]
\begin{center}
\includegraphics[
height=1.8455in,
width=5.9205in
]%
{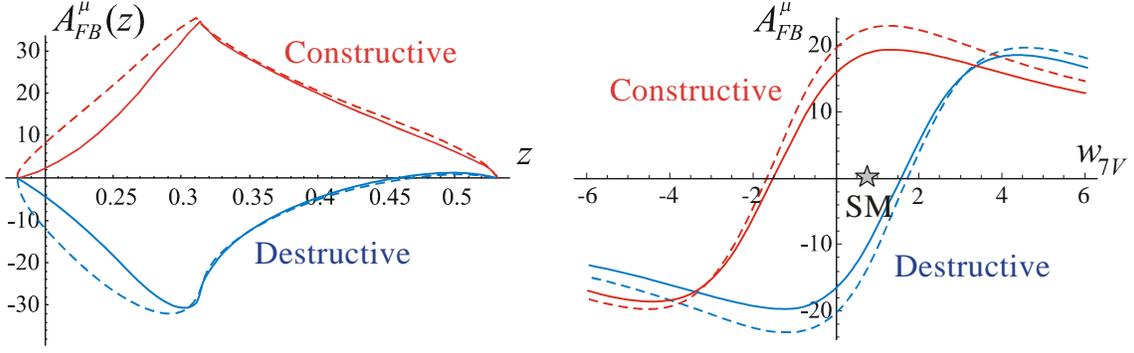}%
\caption{Left: $A_{FB}^{\mu}$ (in \%) as a function of $z$, in the SM. Right:
Integrated $A_{FB}^{\mu}$ (in \%) as a function of $w_{7V}$, with $w_{7A}$
fixed at its SM value. Red (blue) lines correspond to constructive
(destructive) interference between ICPV and vector current contributions,
while plain\ (dashed) lines correspond to $a_{1}^{P}\left(  z\right)
_{Dalitz}$ ($a_{1}^{P}\left(  z\right)  _{ChPT}$), respectively.}%
\label{Fig3}%
\end{center}
\end{figure}

\paragraph{\textit{Forward-backward asymmetry}:}

The differential forward-backward asymmetry\cite{2pp,AFB} is defined by%
\begin{equation}
A_{FB}^{\ell}\left(  z\right)  =\frac{%
{\displaystyle\int_{0}^{y_{0}}}
\dfrac{d^{2}\Gamma}{dydz}dy-%
{\displaystyle\int_{-y_{0}}^{0}}
\dfrac{d^{2}\Gamma}{dydz}dy}{%
{\displaystyle\int_{0}^{y_{0}}}
\dfrac{d^{2}\Gamma}{dydz}dy+%
{\displaystyle\int_{-y_{0}}^{0}}
\dfrac{d^{2}\Gamma}{dydz}dy}\;, \label{Eq12}%
\end{equation}
with%
\begin{equation}
y=\frac{P\cdot(p-p^{\prime})}{M_{K}^{2}},\;\;y_{0}=\frac{1}{2}\beta_{\ell
}\beta_{\pi}\;,
\end{equation}
and $\beta_{\ell}^{2}=1-4r_{\ell}^{2}/z$, $\beta_{\pi}^{2}=\lambda\left(
1,r_{\pi}^{2},z\right)  $, $\lambda\left(  a,b,c\right)  =a^{2}+b^{2}%
+c^{2}-2\left(  ab+bc+ac\right)  .$ The variable $y$ is related to the angle
between the $K$ and $\ell^{-}$ momenta in the dilepton rest-frame (hence the
name forward-backward), and also, by definition, to the energy difference
$E_{\ell^{-}}-E_{\ell^{+}}$ in the $K$ rest-frame (one then speaks of lepton
energy asymmetry).

This observable requires CP-violation, and arises from the $\operatorname{Re}%
[\mathcal{M}_{\mathrm{1}^{--}}^{\ast}\left(  \mathcal{M}_{\mathrm{0}^{++}%
}+\mathcal{M}_{\mathrm{2}^{++}}\right)  ]$ interference term. Let us start by
assuming that $\mathcal{M}_{\mathrm{2}^{++}}$ is negligible, as for the total
rates. Then:%
\begin{equation}
A_{FB}^{\ell}\left(  z\right)  =\frac{A_{1}^{\ell}\left(  z\right)  }%
{d\Gamma/dz}\operatorname{Re}\left(  \varepsilon^{\ast}W_{S}^{\ast}\left(
z\right)  \mathcal{F}_{\ell}\left(  z\right)  \right)  +\frac{A_{2}^{\ell
}\left(  z\right)  }{d\Gamma/dz}\operatorname{Im}\left(  \lambda_{t}%
y_{7V}\right)  \operatorname{Im}\mathcal{F}_{\ell}\left(  z\right)  \;,
\label{Eq12b}%
\end{equation}
with $A_{1,2}^{\ell}\left(  z\right)  $ some combinations of constants and
form-factors. The electronic asymmetry $A_{FB}^{e}\left(  z\right)  $ is
negligible in this case since $\mathcal{F}_{e}\left(  z\right)  $ is helicity
suppressed, while for the muon it is shown in Fig.\ref{Fig3} in the case of
the Standard Model. Note that the theoretical control gained over this
quantity would be difficult to improve since it relies on the specific
parametrization of $\mathcal{M}_{\mathrm{\gamma\gamma}}$. In addition, the
experimental sensitivity required to measure $A_{FB}^{\ell}\left(  z\right)  $
is unlikely to be achieved soon. We will therefore not consider $A_{FB}^{\ell
}\left(  z\right)  $ anymore, but rather concentrate on the integrated
asymmetry:%
\begin{equation}
A_{FB}^{\ell}=\frac{%
{\displaystyle\int_{4r_{\ell}^{2}}^{(1-r_{\pi})^{2}}}
dz\left(
{\displaystyle\int_{0}^{y_{0}}}
\dfrac{d^{2}\Gamma}{dydz}dy-%
{\displaystyle\int_{-y_{0}}^{0}}
\dfrac{d^{2}\Gamma}{dydz}dy\right)  }{%
{\displaystyle\int_{4r_{\ell}^{2}}^{(1-r_{\pi})^{2}}}
dz\left(
{\displaystyle\int_{0}^{y_{0}}}
\dfrac{d^{2}\Gamma}{dydz}dy+%
{\displaystyle\int_{-y_{0}}^{0}}
\dfrac{d^{2}\Gamma}{dydz}dy\right)  }=\frac{N\left(  E_{\ell^{-}}>E_{\ell^{+}%
}\right)  -N\left(  E_{\ell^{-}}<E_{\ell^{+}}\right)  }{N\left(  E_{\ell^{-}%
}>E_{\ell^{+}}\right)  +N\left(  E_{\ell^{-}}<E_{\ell^{+}}\right)  }\;.
\label{Eq12c}%
\end{equation}
Compared to $A_{FB}^{\mu}\left(  z\right)  $, it is more stable:%
\begin{equation}
A_{FB}^{\mu}=\left(  1.3\left(  1\right)  w_{7V}\pm1.7(2)\left|  a_{S}\right|
\right)  \cdot10^{-12}\,\text{\thinspace}/\,\mathcal{B}_{\mathrm{V,A}}%
^{\mu^{+}\mu^{-}}\;. \label{Eq13}%
\end{equation}
In the Standard Model, $A_{FB}^{\mu}=\left(  20\pm4\right)  \%$ for
constructive and $\left(  -12\pm4\right)  \%$ for destructive interference,
with the error coming from varying $a_{1}^{\pi}\left(  z\right)  $ between
$\mathcal{O}\left(  p^{4}\right)  $ ChPT and Dalitz accounting for $\pm2\%$.

For general axial currents, it is clear that $A_{FB}^{\mu}$ decreases when
$w_{7A}$ increases since it does not contribute to the interference
Eq.(\ref{Eq12b}). For $w_{7V}$, the interference is linear while the rate is
quadratic, and thus $|A_{FB}^{\mu}|$ reaches a maximum, around $23\%$ for
$w_{7A}$ at its SM value, before decreasing again, see Fig.\ref{Fig3}. The
absolute maximum for $A_{FB}^{\mu}$ is around $\pm25\%$ for $w_{7A}\approx0$
and $w_{7V}\approx\pm1$ or $\pm4$, depending on the direct - indirect CPV
interference sign.

In the SM, even if $A_{FB}^{\mu}$ is polluted by the theoretical error on the
two-photon amplitude, its measurement could fix the sign of $a_{S}$. As can be
seen in Fig.\ref{Fig3}, this remains true if New Physics is found from the
measurements of $K_{L}\rightarrow\pi^{0}\ell^{+}\ell^{-}$ total rates with
$|w_{7V}|\lesssim2$.

Let us now consider the interference term $\operatorname{Re}[\mathcal{M}%
_{\mathrm{1}^{--}}^{\ast}\mathcal{M}_{\mathrm{2}^{++}}]$. The amplitude
$\mathcal{M}_{\mathrm{2}^{++}}$ is discussed in detail in Ref.\cite{BDI03}. It
leads to a helicity-allowed, but phase-space suppressed contribution to
$A_{FB}^{\ell}$, hence contributes mostly for the electronic mode.
Unfortunately, the theoretical control on the $\mathcal{M}_{\mathrm{2}^{++}}$
amplitude is not good as it depends on unknown phenomenological parameters.
Though sufficient for deriving, from $K_{L}\rightarrow\pi^{0}\gamma\gamma$, a
tight upper bound on the $\mathrm{2}^{++}$ contribution to $C_{\gamma\gamma
}^{\ell}$ in Eq.(\ref{Eq11})\cite{BDI03}, $A_{FB}^{e}$ remains largely
unconstrained. Indeed, both $\mathcal{M}_{\mathrm{1}^{--}}^{\ast}$ and
$\mathcal{M}_{\mathrm{2}^{++}}$ contribute mostly at low $z$, and their
sizeable interference can generate $A_{FB}^{e}$ anywhere between $0\%$ and
about $\pm60\%$\cite{2pp}, depending on the phenomenological parameters of
Ref.\cite{BDI03}.

Concerning $A_{FB}^{\mu}$, the situation is better. The impact of
$\mathcal{M}_{\mathrm{2}^{++}}$ corresponds to an additional $\pm3\%$
uncertainty, and therefore does not affect the potential of $A_{FB}^{\mu}$ in
determining the sign of $a_{S}$.

\section{$K_{L}\rightarrow\pi^{0}\ell^{+}\ell^{-}$ with generic new physics operators}

In addition to the modification of vector and axial-vector couplings
considered in the previous section, new four-fermion effective interactions
could be generated by the integration of New Physics heavy degrees of freedom.
The effective Hamiltonian comprising all the possible dimension-six
semi-leptonic four-fermion structures\cite{Michel} relevant for $K_{L}%
\rightarrow\pi^{0}\ell^{+}\ell^{-}$ (as well as quark bilinear electromagnetic
couplings) reads%
\begin{equation}
\mathcal{H}_{eff}=\mathcal{H}_{eff}^{\mathrm{V,A}}+\mathcal{H}_{eff}%
^{\mathrm{P,S}}+\mathcal{H}_{eff}^{\mathrm{T,\tilde{T}}}+\mathcal{H}%
_{eff}^{\mathrm{EMO}}\;, \label{NP0}%
\end{equation}
with $\mathcal{H}_{eff}^{\mathrm{V,A}}$ given in Eq.(\ref{Eq2}) and%
\begin{align}
\mathcal{H}_{eff}^{\mathrm{P,S}}  &  =\frac{G_{F}^{2}M_{W}^{2}}{\pi^{2}%
}\frac{m_{s}m_{\ell}}{M_{W}^{2}}\left[  y_{P}\left(  \bar{s}d\right)  \left(
\bar{\ell}\gamma_{5}\ell\right)  +y_{S}\left(  \bar{s}d\right)  \left(
\bar{\ell}\ell\right)  \right]  +h.c.\label{NP1}\\
\mathcal{H}_{eff}^{\mathrm{T,\tilde{T}}}  &  =\frac{G_{F}^{2}M_{W}^{2}}%
{\pi^{2}}\frac{m_{s}m_{\ell}}{M_{W}^{2}}\left[  y_{T}\left(  \bar{s}%
\sigma_{\mu\nu}d\right)  \left(  \bar{\ell}\sigma^{\mu\nu}\ell\right)
+y_{\tilde{T}}\left(  \bar{s}\sigma_{\mu\nu}d\right)  \left(  \bar{\ell}%
\sigma^{\mu\nu}\gamma_{5}\ell\right)  \right]  +h.c.\label{NP2}\\
\mathcal{H}_{eff}^{\mathrm{EMO}}  &  =\frac{G_{F}}{\sqrt{2}}M_{K}\frac{Q_{d}%
e}{16\pi^{2}}\left[  y_{\gamma}^{\pm}\left(  \bar{s}\sigma_{\mu\nu}\left(
1\pm\gamma_{5}\right)  d\right)  F^{\mu\nu}\right]  +h.c.\;\;. \label{NP3}%
\end{align}
A low scale ($\mu\lesssim m_{c}$) is understood for the evaluation of the
Wilson coefficients, quark masses and matrix elements of the operators in the
above equation. Dimension-eight operators, containing two powers of the
external momenta, are not considered as they are very small in the SM and are
expected to remain so in the presence of New Physics (see discussion in
\cite{BBKU02}).

Our goal is to analyze the impact of these new operators on the $K_{L}%
\rightarrow\pi^{0}\ell^{+}\ell^{-}$ branching fractions and asymmetries in a
model-independent way. Still, some comments on specific scenarios behind the
various operators are in order:

\begin{enumerate}
\item For the scalar and pseudoscalar operators $Q_{S}=\left(  \bar
{s}d\right)  \left(  \bar{\ell}\ell\right)  $ and $Q_{P}=\left(  \bar
{s}d\right)  \left(  \bar{\ell}\gamma_{5}\ell\right)  $, we have explicitly
included the $m_{s}m_{\ell}/M_{W}^{2}$ helicity suppression factor to give a
realistic description of models where these operators are generated from an
extended Higgs sector. For example, large $y_{P,S}$ can arise in the MSSM with
large $\tan\beta$ (see e.g. \cite{TanBeta,Foster05}) and sizeable trilinear
soft-breaking couplings. This kind of scenarios has been analyzed in many
works, but usually focuses on other decay modes. See e.g.
Refs.\cite{IsidoriRetico,BBKU02} for a MSSM analysis with emphasis on the
$K,B\rightarrow\ell^{+}\ell^{-}$ decays.

\item The tensor and pseudotensor operators $Q_{T}=\left(  \bar{s}\sigma
_{\mu\nu}d\right)  \left(  \bar{\ell}\sigma^{\mu\nu}\ell\right)  $ and
$Q_{\tilde{T}}=\left(  \bar{s}\sigma_{\mu\nu}d\right)  \left(  \bar{\ell
}\sigma^{\mu\nu}\gamma_{5}\ell\right)  =i\varepsilon^{\mu\nu\rho\sigma}\left(
\bar{s}\sigma_{\mu\nu}d\right)  \left(  \bar{\ell}\sigma_{\rho\sigma}%
\gamma_{5}\ell\right)  $, to our knowledge, have not been included in studies
of $K_{L}\rightarrow\pi^{0}\ell^{+}\ell^{-}$ so far. These modes are however
the most promising source of information on $Q_{T,\tilde{T}}$, since these
cannot contribute to $K\rightarrow\ell^{+}\ell^{-}$. Though they do not arise
in the SM, they do in the MSSM but, in addition to being helicity suppressed,
they are usually suppressed by loop factors\cite{BBKU02}. Similar operators
have been considered in $\Delta F=2$ processes (see e.g.
Refs.\cite{Ciuchini98,DF2}) and for the $B\rightarrow X\ell^{+}\ell^{-}$ rate
and asymmetries (see e.g. Ref.\cite{Btensor}).

\item For completeness, we have included the dimension-five electromagnetic
tensor operators $Q_{\gamma}^{\pm}=\left(  \bar{s}\sigma_{\mu\nu}\left(
1\pm\gamma_{5}\right)  d\right)  F^{\mu\nu}$. These were considered for
example in Refs.\cite{BurasCIRS99,DAmbrosioGao}. Since the $\sigma_{\mu\nu
}\gamma_{5}$ part does not contribute to $K_{L}\rightarrow\pi^{0}\ell^{+}%
\ell^{-}$, only $y_{\gamma}\equiv y_{\gamma}^{+}+y_{\gamma}^{-}$ is accessible
here. Note that in principle these operators already arise in the Standard
Model, however they are too small to affect $K_{L}\rightarrow\pi^{0}\ell
^{+}\ell^{-}$. In the MSSM, they are correlated with the chromomagnetic tensor
operators and thus strongly constrained by other
observables\cite{BurasCIRS99,DAmbrosioGao}.

\item In the last section, we consider the general framework in which neither
$Q_{S,P}$ nor $Q_{T,\tilde{T}}$ are helicity-suppressed, i.e. we remove the
$m_{s}m_{\ell}/M_{W}^{2}$ factors in Eqs.(\ref{NP1},\ref{NP2}). A large class
of models with such helicity-allowed FCNC operators are theories with
leptoquark interactions (for a review, see \cite{DavidsonBC94}), among which
specific GUT models. Alternatively, SUSY without R-parity can also induce
helicity-allowed $Q_{S,P}$ interactions through tree-level sneutrino exchanges
(see e.g. \cite{Rparity}).
\end{enumerate}

The distinction between helicity-suppressed and helicity-allowed scenarios is
necessary as the corresponding signatures, i.e. impacts on $K_{L}%
\rightarrow\pi^{0}e^{+}e^{-}$ and $K_{L}\rightarrow\pi^{0}\mu^{+}\mu^{-}$,
will obviously be very different. Let us now analyze these impacts systematically.

\subsection{Scalar and pseudoscalar operators}

The relevant matrix element reads%
\begin{equation}
\langle\pi^{0}|\bar{s}d|K^{0}\rangle=-\frac{M_{K}^{2}-M_{\pi}^{2}}{\sqrt
{2}\left(  m_{s}-m_{d}\right)  }f_{0}\left(  z\right)  \; \label{SP1}%
\end{equation}
in the sign convention of Eq.(\ref{Eq4}). This matrix element is enhanced
compared to its vector counterpart due to the large value of the quark
condensate (i.e., the large ratio of meson masses over quark masses).

The scalar (pseudoscalar) operator produces the lepton pair in a CP-even
$0^{++}$ (CP-odd $0^{-+}$) state, therefore it is the real (imaginary) part of
its Wilson coefficient that contributes to $K_{L}\rightarrow\pi^{0}\ell
^{+}\ell^{-}$:%
\begin{align}
\mathcal{M}_{\mathrm{P}}  &  =\frac{G_{F}^{2}M_{W}^{2}}{\pi^{2}}\frac{m_{\ell
}}{M_{W}^{2}}\left(  M_{K}^{2}-M_{\pi}^{2}\right)  \;i\operatorname{Im}%
y_{P}\;f_{0}\left(  z\right)  \;\left\{  \bar{u}_{p}\gamma_{5}v_{p^{\prime}%
}\right\}  \;,\label{SP2}\\
\mathcal{M}_{\mathrm{S}}  &  =\frac{G_{F}^{2}M_{W}^{2}}{\pi^{2}}\frac{m_{\ell
}}{M_{W}^{2}}\left(  M_{K}^{2}-M_{\pi}^{2}\right)  \;\operatorname{Re}%
y_{S}\;f_{0}\left(  z\right)  \;\left\{  \bar{u}_{p}v_{p^{\prime}}\right\}
\;. \label{SP3}%
\end{align}
To reach these expressions, $m_{d}$ has been neglected against $m_{s}$ in
Eq.(\ref{SP1}). The pseudoscalar current interferes with the
helicity-suppressed pseudoscalar part of the axial-vector current, while the
scalar current interferes with the helicity-suppressed two--photon 0$^{++}$
contribution. Since in addition $Q_{S}$ and $Q_{P}$ are themselves
helicity-suppressed, only the muon mode can be affected by $y_{S,P}$.

\paragraph{\textit{For the pseudoscalar operator,}}

including also the $\mathrm{V}$ and $\mathrm{A}$ contributions, the
differential rate reads:%
\begin{gather}
\frac{d\Gamma_{\mathrm{V,A\&P}}}{dz}=\frac{G_{F}^{2}M_{K}^{5}\alpha^{2}}%
{64\pi^{3}}\left(  \operatorname{Im}\lambda_{t}\right)  ^{2}\beta_{\ell}%
\beta_{\pi}\left(  \frac{\beta_{\pi}^{2}}{6}A_{+}\left(  f_{+}\left(
z\right)  \right)  ^{2}+A_{0}\left(  f_{0}\left(  z\right)  \right)
^{2}\right)  ,\label{SP4}\\
A_{+}=w_{7A}^{2}\beta_{\ell}^{2}+w_{7V}^{2}\frac{3-\beta_{\ell}^{2}}%
{2},\;\;A_{0}=\frac{r_{\ell}^{2}}{z}\left(  w_{7A}\left(  1-r_{\pi}%
^{2}\right)  +z\rho_{P}\operatorname{Im}y_{P}\right)  ^{2},\nonumber
\end{gather}
with the prefactor%
\begin{equation}
\rho_{P}=\frac{1}{\operatorname{Im}\lambda_{t}}\frac{1}{2\pi\sin^{2}\theta
_{W}}\frac{M_{K}^{2}-M_{\pi}^{2}}{M_{W}^{2}}\approx0.18\;. \label{SP5}%
\end{equation}
$\operatorname{Im}y_{P}\sim\mathcal{O}\left(  10\right)  $ is thus required to
get $\mathcal{O}\left(  1\right)  $ effects. Numerically, the contributions to
the total rates, to be added to Eq.(\ref{Eq11}), are
\begin{subequations}
\label{SP6}%
\begin{align}
\mathcal{B}_{\mathrm{P}}^{e^{+}e^{-}}  &  =\left(  1.9\,w_{7A}%
\,\operatorname{Im}y_{P}+0.038\left(  \operatorname{Im}y_{P}\right)
^{2}\right)  \cdot10^{-17}\;,\\
\mathcal{B}_{\mathrm{P}}^{\mu^{+}\mu^{-}}  &  =\left(  0.26\,w_{7A}%
\,\operatorname{Im}y_{P}+0.0085\left(  \operatorname{Im}y_{P}\right)
^{2}\right)  \cdot10^{-12}\;,
\end{align}
showing the very strong helicity suppression at play for the electron mode.

\paragraph{\textit{For the scalar operator,}}

from Eqs.(\ref{Eq8b}) and (\ref{SP3}), one immediately gets for the total
$0^{++}$ contribution:%
\end{subequations}
\begin{equation}
\frac{d\Gamma_{\mathrm{\gamma\gamma\&S}}}{dz}=\frac{G_{8}^{2}M_{K}^{5}%
\alpha_{em}^{4}}{512\pi^{7}}\beta_{\pi}\beta_{\ell}^{3}\frac{r_{\ell}^{2}}%
{z}\left|  \mathcal{F}_{\ell}\left(  z\right)  -\rho_{S}\operatorname{Re}%
y_{S}\,z\,f_{0}\left(  z\right)  \right|  ^{2}\;, \label{SP8}%
\end{equation}
with the suppression factor%
\begin{equation}
\rho_{S}=-\frac{\sqrt{2}\pi}{\sin^{2}\theta_{W}}\frac{\alpha G_{F}}%
{\alpha_{em}^{2}G_{8}}\frac{M_{K}^{2}-M_{\pi}^{2}}{M_{W}^{2}}\approx0.13\;.
\label{SP9}%
\end{equation}
Since $\mathcal{F}_{\mu}\left(  z\right)  \sim\mathcal{O}\left(  1\right)  $,
the helicity suppression $M_{W}^{-2}$ turns out to be nearly compensated.
Performing the $z$ integral, we find
\begin{subequations}
\label{SP10}%
\begin{align}
\mathcal{B}_{\mathrm{S}}^{e^{+}e^{-}}  &  =(1.5\left(  3\right)
\operatorname{Re}y_{S}+0.0039\left(  \operatorname{Re}y_{S}\right)  ^{2}%
)\cdot10^{-16}\;,\\
\mathcal{B}_{\mathrm{S}}^{\mu^{+}\mu^{-}}  &  =(0.04\left(  1\right)
\operatorname{Re}y_{S}+0.0041\left(  \operatorname{Re}y_{S}\right)  ^{2}%
)\cdot10^{-12}\;.
\end{align}
The error on the interference term is estimated by varying the distribution
$a_{1}^{\pi}\left(  z\right)  $ between $\mathcal{O}\left(  p^{4}\right)  $
ChPT and Dalitz (giving respectively $0.034$ and $0.053$ for the muonic mode).

\paragraph{\textit{Total rates:}}

The $Q_{S}$ and $Q_{P}$ operators do not affect the electronic mode due to
their strong helicity suppression. For the muonic mode, combining
Eqs.(\ref{SP6},\ref{SP10}) with Eq.(\ref{Eq11}) and fixing $w_{7A,7V}$ at
their SM values Eq.(\ref{Eq3}), the impacts on the total rate are summarized
in the first three columns of Table \ref{TableSP}.%

\begin{table}[t] \centering
$%
\begin{tabular}
[c]{cccccc}\hline
& Enhancement & Enhancement & Maximal & Bound from & Experimental\\
& of 50\% & of 100\% & suppression & $K_{L}\rightarrow\mu^{+}\mu^{-}$ & bound
(Eq.(\ref{Eq11c}))\\\hline
$\operatorname{Im}y_{P}$ & $-20,40$ & $-30,50$ & 7\% for $\operatorname{Im}%
y_{P}\approx10\;\,\,$ & $\left|  \operatorname{Im}y_{P}\right|  \lesssim8$ &
$|\operatorname{Im}y_{P}|\lesssim220$\\
$\operatorname{Re}y_{S}$ & $-45,35$ & $-65,55$ & 1\% for $\operatorname{Re}%
y_{S}\approx-5$ & $\left|  \operatorname{Re}y_{S}\right|  \lesssim50$ &
$|\operatorname{Re}y_{S}|\lesssim300$\\\hline
\end{tabular}
\ $%
\caption{Numerical analysis of scalar and pseudoscalar
operator impacts on $\mathcal{B}%
^{\mu^+\mu^-}$.
\label{TableSP}}
\end{table}%

A well-motivated scenario in which $\operatorname{Re}y_{S}$ and
$\operatorname{Im}y_{P}$ can be large is the MSSM for large values of
$\tan\beta$\cite{TanBeta,Foster05}. In that context, the contributions of
$Q_{S,P}$ are related to those of $Q_{P}^{\prime}=\left(  \bar{s}\gamma
_{5}d\right)  \left(  \bar{\ell}\gamma_{5}\ell\right)  $ and $Q_{S}^{\prime
}=(\bar{s}\gamma_{5}d)(\bar{\ell}\ell)$: $y_{S,P}=y_{P,S}^{\prime}$, with
$y_{P}^{\prime}$ and $y_{S}^{\prime}$ further correlated. The contributions of
$Q_{P,S}^{\prime}$ to $K_{L}\rightarrow\mu^{+}\mu^{-}$ were analyzed for
example in Ref.\cite{IsidoriRetico}, with the result that values of a few tens
for $y_{S,P}^{\prime}$ are compatible with $\Delta S=2$ and $B$-physics
data\footnote{Assuming a degenerate SUSY spectrum $M_{\tilde{d}}\sim\left|
\mu\right|  \sim M_{A}\sim M_{\tilde{g}}$, the gluino contributions to
$y_{S,P}$ scale as\cite{IsidoriRetico}%
\[
y_{S,P}\sim\left(  M_{W}^{2}/M_{A}^{2}\right)  \tan^{3}\beta\left(
1+0.01\tan\beta\operatorname{sign}\mu\right)  ^{-2}\left(  \left(  \delta
_{LL}^{D}\right)  _{12}+18\left(  \delta_{RR}^{D}\right)  _{13}\left(
\delta_{LL}^{D}\right)  _{32}\right)  \;.
\]
With $\tan\beta\sim50$, $\mu>0$ (as favored from the muon $g-2$) and $M_{A}$
in the range $300-500$ GeV\cite{Foster05}, one gets $y_{S,P}\sim
\mathcal{O}\left(  15\right)  $ with the single mass-insertion $(\delta
_{LL(RR)}^{D})_{12}\sim10^{-2}$ (compatible with $\varepsilon_{K}$ constraints
\cite{Ciuchini98}) and $y_{S,P}\sim\mathcal{O}\left(  30\right)  $ with the
double one $\left(  \delta_{RR}^{D}\right)  _{13}\left(  \delta_{LL}%
^{D}\right)  _{32}\sim10^{-3}$ (constrained by $\Delta M_{s,d}$
\cite{Foster05}).}.

Without restricting ourselves to the MSSM, we can investigate the constraints
on $y_{S,P}$ derived from the experimental $K_{L}\rightarrow\mu^{+}\mu^{-}$
rate under the assumption that $y_{S,P}=y_{P,S}^{\prime}$ approximately holds.
This analysis is presented in Appendix A, leading to $\left|
\operatorname{Im}y_{S}^{\prime}\right|  \lesssim8$ and $\left|
\operatorname{Re}y_{P}^{\prime}\right|  \lesssim35$ for $y_{7A}^{\prime
}=y_{7A}=-0.68$ (for the $Q_{7A}^{\prime}=\left(  \bar{s}\gamma_{\mu}%
\gamma_{5}d\right)  \left(  \bar{\ell}\gamma^{\mu}\gamma_{5}\ell\right)  $
operator). Allowing for New Physics in the axial-vector current with the bound
$\left|  y_{7A}^{\prime}\right|  \lesssim3$ corresponds to $\left|
\operatorname{Re}y_{P}^{\prime}\right|  \lesssim70$, a much larger range since
these two contributions interfere in the rate.

Our numerical analysis can be summarized drawing the allowed regions on the
$\mathcal{B}^{e^{+}e^{-}}$--$\mathcal{B}^{\mu^{+}\mu^{-}}$ plane
(Fig.\ref{Fig4}). Compared to the region spanned for general values of
$y_{7A,7V}$ (Fig.\ref{Fig2}), turning on $y_{S,P}$ basically extends the
vertical spread since only the muon mode is affected, and this mostly in the
upward direction (i.e., enhancements). This is illustrated in Fig.\ref{Fig4}
by the light-blue/dark-blue region, corresponding to $|\operatorname{Re}%
y_{S}|<90$ / $|\operatorname{Im}y_{P}|<35$, respectively. Imposing further the
constraints from $K_{L}\rightarrow\mu^{+}\mu^{-}$ under the assumption
$y_{7A,S,P}=y_{7A,P,S}^{\prime}$ gives the yellow region. Obviously, the
sensitivity of $K_{L}\rightarrow\pi^{0}\mu^{+}\mu^{-}$ to $Q_{S,P}$ is quite good.

It should also be noted that $Q_{S,P}$ contribute mostly for large $z$. While,
as explained in Ref.\cite{IsidoriSmithUnter}, introducing a cut off at
$z\approx4r_{\ell}^{2}$ can reduce the contribution of the two-photon
amplitude with respect to the $Q_{7A,7V}$ ones, thereby reducing the
theoretical error, such a procedure would also reduce the sensitivity to
$\operatorname{Im}y_{P}$ and $\operatorname{Re}y_{S}$ significantly, and may
thus not be desirable.%

\begin{figure}
[ptb]
\begin{center}
\includegraphics[
height=2.8228in,
width=6.4835in
]%
{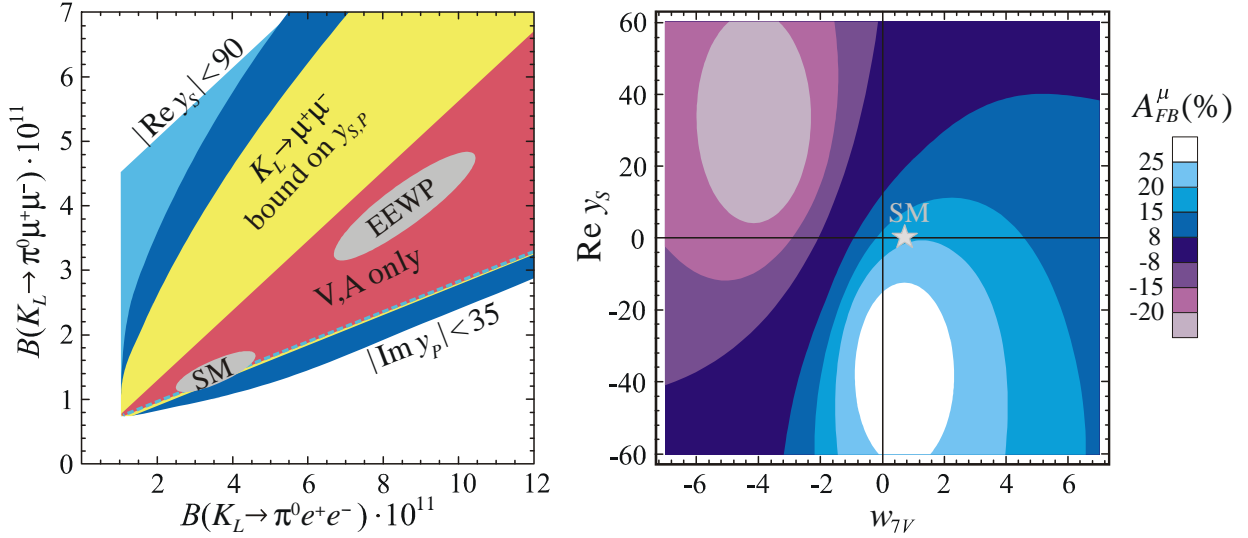}%
\caption{Left: Impacts of scalar and pseudoscalar operators in the
$\mathcal{B}^{e^{+}e^{-}}-\mathcal{B}^{\mu^{+}\mu^{-}}$ plane of
Fig.\ref{Fig2}. Light blue (dark blue) corresponds to arbitrary $y_{7A,7V}$
together with $\operatorname{Re}y_{S}<90$ ( $\operatorname{Im}y_{P}<35$),
resp., while the yellow region corresponds to $y_{7V,7A,S,P}$ arbitrary but
compatible with $\mathcal{B}\left(  K_{L}\rightarrow\mu^{+}\mu^{-}\right)
^{\exp}$ (see text). The dashed, light-blue line indicates the lower extent of
the corresponding region. Right: The asymmetry $A_{FB}^{\mu}$ as a function of
$w_{7V}$ and $\operatorname{Re}y_{S}$, assuming constructive interference
($a_{S}<0$), $y_{7A}=-0.68$ and $\operatorname{Im}y_{P}=0$.}%
\label{Fig4}%
\end{center}
\end{figure}

\paragraph{\textit{Forward-backward asymmetry}:}

$\operatorname{Re}y_{S}$ enters the numerator of $A_{FB}^{\mu}$ through the
interference term $\mathcal{M}_{\mathrm{ICPV}}^{\ast}\mathcal{M}_{\mathrm{S}}$
only:
\end{subequations}
\begin{equation}
A_{FB}^{\mu}=(1.3\left(  1\right)  w_{7V}\pm1.7(2)\left|  a_{S}\right|
\mp0.057\left|  a_{S}\right|  \operatorname{Re}y_{S})\cdot10^{-12}%
\,\text{\thinspace}/\,\mathcal{B}_{\mathrm{V,A,S,P}}^{\mu^{+}\mu^{-}}\;.
\label{SP11}%
\end{equation}
There is no interference between $\mathcal{M}_{\mathrm{S}}$ and $\mathcal{M}%
_{\mathrm{V}}$ because of their 90$%
{{}^\circ}%
$ relative phase. For this reason, $A_{FB}^{\mu}$ goes to zero when $y_{7V}$
and/or $\operatorname{Re}y_{S}$ becomes large, and reaches its maximum for
moderate values, as shown in Fig.\ref{Fig4} for $y_{7A}=-0.68$ and
$\operatorname{Im}y_{P}=0$. If these latter two values are enhanced, since
they contribute only to $\mathcal{B}_{\mathrm{V,A,S,P}}^{\mu^{+}\mu^{-}}$,
$A_{FB}^{\mu}$ decreases, i.e. the figure remains the same but the absolute
size of $A_{FB}^{\mu}$ is reduced. Finally, the figure for destructive
interference is readily obtained by performing a vertical axis reflexion
($w_{7V}\rightarrow-w_{7V}$) followed by an overall sign change for
$A_{FB}^{\mu}$.

No matter the New Physics behind $y_{7A,7V,S,P}$, $|A_{FB}^{\mu}|$ is always
smaller than $30\%$, i.e., not very far from its SM value Eq.(\ref{Eq13}).
Given the theoretical errors, $A_{FB}^{\mu}$ does not appear very promising to
get a clear signal of New Physics. Nevertheless, as said before, it offers a
very interesting possibility of constraining the relative signs of
$\operatorname{Re}y_{S}$, $y_{7V}$ and $a_{S}$ when considered in conjunction
with the $K_{L}\rightarrow\pi^{0}\ell^{+}\ell^{-}$ total rates.

\subsection{Tensor and pseudo-tensor operators}

Let us now turn to the tensor operators of Eqs.(\ref{NP2}) and (\ref{NP3}).
The relevant matrix element assumes the form%
\begin{equation}
\langle\pi^{0}\left(  K\right)  |\bar{s}\sigma^{\mu\nu}d|K^{0}\left(
P\right)  \rangle=i\frac{P^{\mu}K^{\nu}-P^{\nu}K^{\mu}}{\sqrt{2}M_{K}}%
B_{T}\left(  z\right)  , \label{NPT1}%
\end{equation}
with $\langle\pi^{0}|\bar{s}\sigma^{\mu\nu}\gamma_{5}d|K^{0}\rangle$ obtained
through $\sigma^{\mu\nu}\gamma_{5}=i\varepsilon^{\mu\nu\rho\sigma}\sigma
_{\rho\sigma}$. The tensor form-factor was studied on the lattice
\cite{Mescia00}, with the result $B_{T}\left(  z\right)  \approx
1.2f_{+}\left(  0\right)  /(1-0.29z)$ at $\mu\simeq m_{c}$ in the
$\overline{MS}$ scheme (an earlier order-of-magnitude estimate may be found in
Ref.\cite{OldBT}).

\paragraph{\textit{The electromagnetic tensor operator}}

produces the lepton pair in a $1^{--}$ state and the transition is
CP-violating:
\begin{equation}
\mathcal{M}_{\mathrm{EMO}}=i\frac{G_{F}\alpha}{\sqrt{2}}\frac{Q_{d}}{2\pi
}\operatorname{Im}y_{\gamma}B_{T}\left(  z\right)  \left\{  \bar{u}_{p}%
\!\not\!%
Pv_{p^{\prime}}\right\}  \;. \label{NPT2}%
\end{equation}
Let us take $\lambda_{T}=\lambda_{+}$, which is good enough for our purpose,
and can be justified in a pole model through the fact that $z<<M_{T,V}%
^{2}/M_{K}^{2}$ with $M_{T,V}$ the nearest vector and tensor resonances. The
effect of $Q_{\gamma}^{\pm}$ can then be absorbed into the vector-current
Wilson coefficient $w_{7V}$\cite{BurasCIRS99}%
\begin{equation}
w_{7V}\rightarrow w_{7V}^{\prime}=w_{7V}+\frac{\operatorname{Im}y_{\gamma}%
}{\operatorname{Im}\lambda_{t}}\frac{Q_{d}}{4\pi}\frac{B_{T}\left(  0\right)
}{f_{+}\left(  0\right)  }\;. \label{NPT3}%
\end{equation}
The above redefinition is independent of the lepton flavor, hence the
$K_{L}\rightarrow\pi^{0}\ell^{+}\ell^{-}$ modes cannot disentangle possible
New Physics effects arising from $Q_{\gamma}^{\pm}$ from those arising in the
vector current electroweak penguins and boxes. Such New Physics effects were
analyzed in Section 2, see Fig.\ref{Fig2}.

\paragraph{\textit{The tensor operator}}

also induces a CP-violating contribution:%
\begin{equation}
\mathcal{M}_{\mathrm{T}}=i\frac{G_{F}\alpha}{\sqrt{2}}\operatorname{Im}%
\lambda_{t}B_{T}\left(  z\right)  r_{\ell}\rho_{T}\operatorname{Im}%
y_{T}\left\{  \bar{u}_{p}\left(  2r_{\ell}%
\!\not\!%
P-\frac{P\cdot(p-p^{\prime})}{M_{K}}\right)  v_{p^{\prime}}\right\}  \;,
\label{NPT4}%
\end{equation}
where we have defined%
\begin{equation}
\rho_{T}=\frac{1}{\operatorname{Im}\lambda_{t}}\frac{2}{\pi\sin^{2}\theta_{W}%
}\frac{m_{s}M_{K}}{M_{W}^{2}}\approx\frac{1}{4} \label{NPT5}%
\end{equation}
for $m_{s}\approx150$ MeV. As for the magnetic operator, the $%
\!\not\!%
P$ part can be absorbed into $w_{7V}$, but now the $m_{\ell}$ dependence
introduces an effective breaking of $\mu-e$ universality in the vector
current:%
\begin{equation}
w_{7V}\rightarrow w_{7V}^{\ell}=w_{7V}+\rho_{T}\operatorname{Im}y_{T}r_{\ell
}^{2}\frac{B_{T}\left(  0\right)  }{f_{+}\left(  0\right)  }\;. \label{NPT6}%
\end{equation}
The second term in Eq.(\ref{NPT4}) is also CP-violating because $P\cdot
(p-p^{\prime})=yM_{K}^{2}$ is CP-odd. It produces the lepton pair again in a
$1^{--}$ state and can thus interfere with both the ICPV and vector operator
contributions, producing an extra contribution to the $A_{+}$ factor defined
in Eq.(\ref{SP4}):%
\begin{equation}
\left(  A_{+}\right)  _{\mathrm{T}}=r_{\ell}^{2}\frac{B_{T}\left(  z\right)
^{2}}{f_{+}\left(  z\right)  ^{2}}\left(  \rho_{T}\operatorname{Im}%
y_{T}\right)  ^{2}\frac{z\beta_{\ell}^{4}}{8}+r_{\ell}^{2}\beta_{\ell}%
^{2}\frac{B_{T}\left(  z\right)  }{f_{+}\left(  z\right)  }\rho_{T}%
\operatorname{Im}y_{T}\left(  w_{7V}^{\ell}-\frac{\alpha_{em}\operatorname{Im}%
\left(  \varepsilon W_{S}\left(  z\right)  \right)  }{\sqrt{8}\pi\alpha
f_{+}\left(  z\right)  \operatorname{Im}\lambda_{t}}\right)  \,. \label{NPT7}%
\end{equation}
In the muonic case, the overall $\beta_{\mu}^{5}$ factor for the first term,
corresponding to an orbital angular momentum of two between the muons ($y$
counts as one unit of angular momentum), makes this contribution significantly
phase-space suppressed compared to the one shifting $w_{7V}$, Eq.(\ref{NPT6}).
Numerically,
\begin{subequations}
\label{NPT8}%
\begin{align}
\mathcal{B}_{\mathrm{T}}^{e^{+}e^{-}}  &  =\left(  \left(  10^{-5}%
+0.08\right)  \left(  \operatorname{Im}y_{T}\right)  ^{2}+\left(  \left(
29+15\right)  w_{7V}\pm\left(  36+18\right)  \left|  a_{S}\right|  \right)
\operatorname{Im}y_{T}\right)  \cdot10^{-19}\;,\\
\mathcal{B}_{\mathrm{T}}^{\mu^{+}\mu^{-}}  &  =\left(  \left(
0.25+0.02\right)  \left(  \operatorname{Im}y_{T}\right)  ^{2}+\left(  \left(
29+4\right)  w_{7V}\pm\left(  36+5\right)  \left|  a_{S}\right|  \right)
\operatorname{Im}y_{T}\right)  \cdot10^{-15}\;,
\end{align}
where the first numbers in each parenthesis come from Eq.(\ref{NPT6}), the
second from Eq.(\ref{NPT7}), and the $\pm$ sign corresponds to $a_{S}%
=\mp\left|  a_{S}\right|  $.

\paragraph{\textit{The pseudotensor operator}}

produces the lepton pair in a $1^{+-}$ state, and is thus CP-conserving%
\end{subequations}
\begin{equation}
\mathcal{M}_{\mathrm{\tilde{T}}}=-\frac{G_{F}\alpha}{\sqrt{2}}%
\operatorname{Im}\lambda_{t}B_{T}\left(  z\right)  r_{\ell}\rho_{T}%
\operatorname{Re}y_{\tilde{T}}\frac{P\cdot(p-p^{\prime})}{M_{K}}\left\{
\bar{u}_{p}\gamma_{5}v_{p^{\prime}}\right\}  \;. \label{NPT10}%
\end{equation}
As none of the other CP-conserving contributions produces such a final state,
it represents a distinct contribution to the rate:%
\begin{equation}
\frac{d\Gamma_{\mathrm{\tilde{T}}}}{dz}=\frac{G_{F}^{2}M_{K}^{5}\alpha^{2}%
}{3072\pi^{3}}\left(  \operatorname{Im}\lambda_{t}\right)  ^{2}\beta_{\ell
}^{3}\beta_{\pi}^{3}r_{\ell}^{2}\left(  \rho_{T}\operatorname{Re}y_{\tilde{T}%
}\right)  ^{2}\left(  B_{T}\left(  z\right)  \right)  ^{2}z\;. \label{NPT11}%
\end{equation}
Numerically, this contribution is phase-space suppressed and very small:
\begin{subequations}
\label{NPT12}%
\begin{align}
\mathcal{B}_{\mathrm{\tilde{T}}}^{e^{+}e^{-}}  &  =7.9\left(
\operatorname{Re}y_{\tilde{T}}\right)  ^{2}\cdot10^{-21}\;,\\
\mathcal{B}_{\mathrm{\tilde{T}}}^{\mu^{+}\mu^{-}}  &  =4.9\left(
\operatorname{Re}y_{\tilde{T}}\right)  ^{2}\cdot10^{-17}\;.
\end{align}

\paragraph{\textit{Total rate and forward-backward asymmetry: }}

Overall, the effects of the $Q_{T,\tilde{T}}$ operators are smaller than those
of $Q_{P,S}$ because of the smaller matrix elements and the phase-space
suppression. In addition, $y_{T,\tilde{T}}<y_{S,P}$ in realistic scenarios,
and tensor operators seem beyond reach. The only exception is the effective
breaking of $\mu-e$ universality in the vector current induced by
$\operatorname{Im}y_{T}$, which slightly extends downwards the ''V,A
only''\ region of Fig.\ref{Fig2} when $\left|  \operatorname{Im}y_{T}\right|
\gtrsim25$.

For $A_{FB}^{\mu}$, $\operatorname{Im}y_{T}$ enters through interference with
$\operatorname{Im}\mathcal{F}_{\mu}\left(  z\right)  $ (not with
$\operatorname{Re}y_{S}$ since they are out of phase by 90$%
{{}^\circ}%
$):%
\end{subequations}
\begin{equation}
A_{FB}^{\mu}=(1.3\left(  1\right)  w_{7V}\pm1.7(2)\left|  a_{S}\right|
\mp0.057\left|  a_{S}\right|  \operatorname{Re}y_{S}+0.033\left(  4\right)
\operatorname{Im}y_{T})\cdot10^{-12}\,\text{\thinspace}/\,\mathcal{B}%
_{\mathrm{V,A,S,P,T,\tilde{T}}}^{\mu^{+}\mu^{-}}. \label{NPT13}%
\end{equation}
$Q_{\tilde{T}}$ does not contribute directly to $A_{FB}^{\mu}$ as it leads to
a real amplitude, out of phase from the $Q_{7A}$ one by 90$%
{{}^\circ}%
$. Therefore, assuming $\operatorname{Im}y_{T}<\operatorname{Re}y_{S}$, no
impact can arise for $A_{FB}^{\mu}$. Similarly, the electronic asymmetry is
not affected, even taking into account interferences with the $\mathcal{M}%
_{\mathrm{2}^{++}}$ piece.

\subsection{Helicity-allowed effective interactions}

We now lift the constraint of helicity suppression in Eqs.(\ref{NP1}) and
(\ref{NP2}), symbolically as:%
\begin{equation}
\frac{G_{F}^{2}M_{W}^{2}}{\pi^{2}}\frac{m_{s}m_{\ell}}{M_{W}^{2}}%
y_{i}\rightarrow\frac{g_{NP}^{2}}{\Lambda_{i}^{2}}\;,\;i=S,P,T,\tilde{T}\;.
\label{EI1}%
\end{equation}
More precisely, all former expressions remain valid provided one makes the
substitutions:
\begin{equation}
\operatorname{Re}y_{S,\tilde{T}},\operatorname{Im}y_{P,T}\rightarrow32\pi
^{2}\times\frac{g_{NP}^{2}}{g^{4}}\times\frac{M_{W}^{2}}{\Lambda_{i}^{2}%
}\times\frac{M_{W}^{2}}{m_{s}m_{\ell}}\rightarrow\left\{
\begin{array}
[c]{l}%
\ell=e:1.7\ 10^{4}/\bar{\Lambda}_{i}^{2}\\
\ell=\mu:83/\bar{\Lambda}_{i}^{2}%
\end{array}
\right.  \;, \label{EI2}%
\end{equation}
where the normalization $\bar{\Lambda}_{i}=(\Lambda_{i}/100\,$TeV$)$
corresponds to typical lower bounds obtained from lepton-number violating
processes (see e.g. \cite{LFV}), and we have assumed $g_{NP}/g^{2}%
\sim\mathcal{O}\left(  1\right)  $. Plugging these expressions in
Eqs.(\ref{SP6},\ref{SP10},\ref{NPT8},\ref{NPT12}), we obtain:
\begin{subequations}
\label{EI3}%
\begin{align}
\mathcal{B}_{\mathrm{S,P,T,\tilde{T}}}^{e^{+}e^{-}}  &  =\left(
\frac{115}{\bar{\Lambda}_{S}^{4}}+\frac{2.6}{\bar{\Lambda}_{S}^{2}}%
+\frac{112}{\bar{\Lambda}_{P}^{4}}+\frac{0.3w_{7A}}{\bar{\Lambda}_{P}^{2}%
}+\frac{2.3}{\bar{\Lambda}_{T}^{4}}+\frac{0.07w_{7V}}{\bar{\Lambda}_{T}^{2}%
}\pm\frac{0.09\left|  a_{S}\right|  }{\bar{\Lambda}_{T}^{2}}+\frac{2.3}%
{\bar{\Lambda}_{\tilde{T}}^{4}}\right)  \cdot10^{-12}\;,\\
\mathcal{B}_{\mathrm{S,P,T,\tilde{T}}}^{\mu^{+}\mu^{-}}  &  =\left(
\frac{29}{\bar{\Lambda}_{S}^{4}}+\frac{4.5}{\bar{\Lambda}_{S}^{2}}%
+\frac{60}{\bar{\Lambda}_{P}^{4}}+\frac{22w_{7A}}{\bar{\Lambda}_{P}^{2}%
}+\frac{1.9}{\bar{\Lambda}_{T}^{4}}+\frac{2.8w_{7V}}{\bar{\Lambda}_{T}^{2}}%
\pm\frac{3.4\left|  a_{S}\right|  }{\bar{\Lambda}_{T}^{2}}+\frac{0.34}%
{\bar{\Lambda}_{\tilde{T}}^{4}}\right)  \cdot10^{-12}\;.
\end{align}
Without helicity-suppression, it is the electronic mode that is the most
sensitive to these operators, simply because of the phase-space suppression in
the muonic mode. Therefore, allowing for these interactions typically produces
total rates in the lower part of the $\mathcal{B}^{e^{+}e^{-}}$--$\mathcal{B}%
^{\mu^{+}\mu^{-}}$ plane, a region which cannot be attained by the New Physics
interactions studied up to now (see Figs.\ref{Fig2} and \ref{Fig4}). Let us
further investigate what signals could be expected.

\paragraph{\textit{Scalar/pseudoscalar operators}:}

As before, assuming that the Wilson coefficients for the scalar and
pseudoscalar operators contributing to $K_{L}\rightarrow\ell^{+}\ell^{-}$ and
$K_{L}\rightarrow\pi^{0}\ell^{+}\ell^{-}$ have approximately the same values,
$\bar{\Lambda}_{S^{\prime},P^{\prime}}\approx\bar{\Lambda}_{P,S}$, the
constraints from $K_{L}\rightarrow\mu^{+}\mu^{-}$ still leave the possibility
of sizeable effects. However, now that helicity suppression is no longer
effective, one gets a very tough constraint from $K_{L}\rightarrow e^{+}e^{-}%
$. From the measurement $\mathcal{B}\left(  K_{L}\rightarrow e^{+}%
e^{-}\right)  ^{\exp}=9_{-4}^{+6}\cdot10^{-12}$\cite{PDG}, and since%
\end{subequations}
\begin{equation}
\mathcal{B}\left(  K_{L}\rightarrow e^{+}e^{-}\right)  _{\mathrm{S,P}}%
\approx6.8\cdot10^{-8}\left(  \frac{1}{\bar{\Lambda}_{S^{\prime}}^{4}%
}+\frac{1}{\bar{\Lambda}_{P^{\prime}}^{4}}\right)  \;, \label{EI4}%
\end{equation}
one gets $\bar{\Lambda}_{S^{\prime}},\bar{\Lambda}_{P^{\prime}}\gtrsim8$. For
such large values, the effect on both $\mathcal{B}^{\ell^{+}\ell^{-}}$ is of a
few percents for $y_{7A}$ at its SM value, hence well beyond reach. Though
$\bar{\Lambda}_{S^{\prime},P^{\prime}}$ and $\bar{\Lambda}_{P,S}$ can be
different in specific models, the large difference needed to get observable
effects on $\mathcal{B}^{\ell^{+}\ell^{-}}$ would require a somewhat
fine-tuned scenario. We can therefore reasonably rule out this possibility.

\paragraph{\textit{Tensor/pseudotensor operators}:}

To get a somehow realistic estimate, we bound them from $K^{+}\rightarrow
\pi^{+}\nu\bar{\nu}$ assuming that the operators $(\bar{s}\sigma^{\mu\nu
}d)(\bar{\nu}\sigma_{\mu\nu}\nu)$ and $(\bar{s}\sigma^{\mu\nu}d)(\bar{\nu
}\sigma_{\mu\nu}\gamma_{5}\nu)$ are governed by the same scale factors
$(g_{NP}/\Lambda_{T,\tilde{T}})^{2}$. Their contribution reads%
\begin{equation}
\mathcal{B}\left(  K^{+}\rightarrow\pi^{+}\nu\bar{\nu}\right)
_{\mathrm{T,\tilde{T}}}\approx4.4\cdot10^{-12}\left(  \frac{1}{\bar{\Lambda
}_{T}^{4}}+\frac{1}{\bar{\Lambda}_{\tilde{T}}^{4}}\right)  \;, \label{EI5}%
\end{equation}
and comparing with Eq.(\ref{EI3}) shows that the sensitivities of
$K\rightarrow\pi\nu\bar{\nu}$ and $K_{L}\rightarrow\pi^{0}\ell^{+}\ell^{-}$ to
tensor interactions are similar.

\begin{figure}
[ptb]
\begin{center}
\includegraphics[
height=3.3166in,
width=3.4013in
]%
{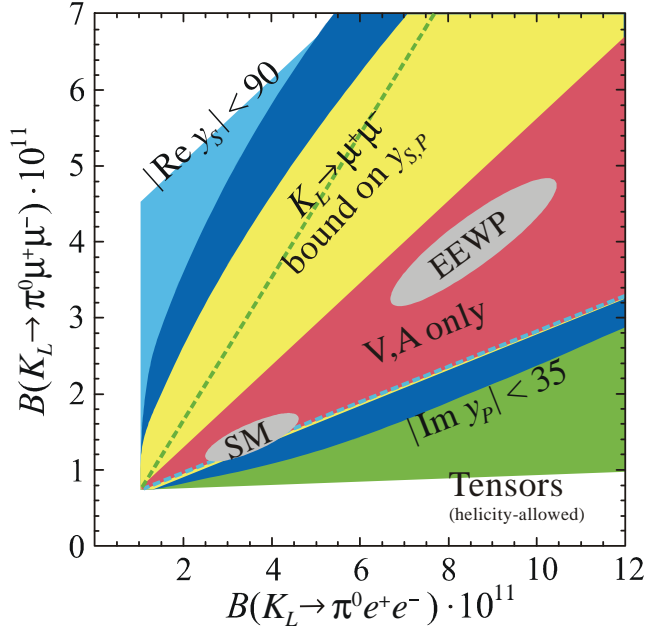}%
\caption{Impact of helicity-allowed tensor and pseudotensor operators in the
$\mathcal{B}^{e^{+}e^{-}}$ -- $\mathcal{B}^{\mu^{+}\mu^{-}}$ plane of
Fig.\ref{Fig4}. The green area (up to the green dashed line) corresponds to
arbitrary $y_{7A,7V}$ together with $\Lambda_{T,\tilde{T}}$ compatible with
$\mathcal{B}\left(  K^{+}\rightarrow\pi^{+}\nu\bar{\nu}\right)  $, as
explained in the text.}%
\label{Fig5}%
\end{center}
\end{figure}

Let us assume that $\bar{\Lambda}_{T}\approx\bar{\Lambda}_{\tilde{T}}$, and
given the SM prediction of $\left(  8.0\pm1.1\right)  \cdot10^{-11}%
$\cite{Buras05} and the current measurement of $14.7_{-8.9}^{+13.0}%
\cdot10^{-11}$\cite{E787E949} (V,A currents do not interfere with tensor
operators for massless neutrinos), one finds $\bar{\Lambda}_{T,\tilde{T}%
}\gtrsim0.35$. Interestingly, for such small values, the charged lepton modes
are quite large, $\mathcal{B}\left(  K_{L}\rightarrow\pi^{0}e^{+}e^{-}\right)
\approx3\cdot10^{-10}$ and $\mathcal{B}\left(  K_{L}\rightarrow\pi^{0}\mu
^{+}\mu^{-}\right)  \approx2\cdot10^{-10}$, i.e., around their current upper
limits Eq.(\ref{Eq11c}). The conclusion of this order-of-magnitude estimate is
thus that there is still room for large effects from tensor operators. The
corresponding allowed region in the $\mathcal{B}^{e^{+}e^{-}}$--$\mathcal{B}%
^{\mu^{+}\mu^{-}}$ plane is shown in Fig.\ref{Fig5}.%

\paragraph{\textit{The forward-backward asymmetry}:}

The muonic asymmetry $A_{FB}^{\mu}$ can be enhanced only by $Q_{S}$ and
$Q_{T}$ contributions, as explained in earlier sections, and this through
interferences with $\mathcal{M}_{\mathrm{ICPV}}^{\ast}$ and $\mathcal{M}%
_{\mathrm{\gamma\gamma-0}^{++}}$, respectively, see Eq.(\ref{NPT13}). Tuning
$y_{7V}$ and the helicity-allowed $1/\bar{\Lambda}_{T}^{2}$ term, a maximum of
about $60\%$ can be reached. Still, this requires large contributions from
$Q_{T}$ with both $K_{L}\rightarrow\pi^{0}\ell^{+}\ell^{-}$ rates above
$10^{-10}$. With $\bar{\Lambda}_{T}\gtrsim1$, this maximum falls to about
$30\%$, i.e. close to the SM value Eq.(\ref{Eq13}).

For the electronic mode, one could think that a small $\bar{\Lambda}_{S}$
could generate a significant asymmetry through the helicity-allowed
interference with $\mathcal{M}_{\mathrm{ICPV}}^{\ast}$. However, this is not
the case because the impact of $Q_{S}$ on $\mathcal{B}^{e^{+}e^{-}}$ is then
much more pronounced. Tensor interactions, for their parts, have a small
impact on $A_{FB}^{e}$ because interference effects with $\mathcal{M}%
_{\mathrm{2}^{++}}$ are helicity-suppressed.

\section{Summary and Conclusion}

The $K_{L}\rightarrow\pi^{0}\ell^{+}\ell^{-}$ modes offer a unique opportunity
to probe a large range of New Physics $\Delta S=1$ effective operators, see
Table \ref{Table1}. They are therefore an essential tool in the investigation
of flavor structures beyond the Standard Model. Our work was to analyze, in a
model-independent way, the possible experimental signatures of these New
Physics interactions.%

\begin{table}[t] \centering
$%
\begin{tabular}
[c]{lcccc}\hline
& $%
\begin{array}
[c]{c}%
\text{Short-distance}\\
\text{operator}%
\end{array}
$ & $%
\begin{array}
[c]{c}%
\text{CP-property}\\
\text{\& }J^{PC}\left(  \bar{\ell}\ell\right)
\end{array}
$ & $%
\begin{array}
[c]{c}%
\text{Helicity-}\\
\text{suppressed}%
\end{array}
$ & $%
\begin{array}
[c]{c}%
\text{Helicity-}\\
\text{allowed}%
\end{array}
$\\\hline
ICPV ($K^{0}$-$\bar{K}^{0}$) & -- & CPV ($1^{--}$) &
\multicolumn{2}{c}{Eq.(\ref{Eq11})}\\
Two-photon & -- & CPC ($0^{++}$) & \multicolumn{2}{c}{Eq.(\ref{Eq11})}\\\hline
Vector & $(\bar{s}\gamma_{\mu}d)(\bar{\ell}\gamma^{\mu}\ell)$ & CPV ($1^{--}%
$) & \multicolumn{2}{c}{Eq.(\ref{Eq11})}\\
Axial-vector & $(\bar{s}\gamma_{\mu}d)(\bar{\ell}\gamma^{\mu}\gamma_{5}\ell)$
& CPV ($0^{-+},1^{++}$) & \multicolumn{2}{c}{Eq.(\ref{Eq11})}\\\hline
Pseudoscalar & $\left(  \bar{s}d\right)  \left(  \bar{\ell}\gamma_{5}%
\ell\right)  $ & CPV ($0^{-+}$) & Eq.(\ref{SP6}) & Eq.(\ref{EI3})\\
Scalar & $\left(  \bar{s}d\right)  \left(  \bar{\ell}\ell\right)  $ & CPC
($0^{++}$) & Eq.(\ref{SP10}) & Eq.(\ref{EI3})\\
Tensor & $\left(  \bar{s}\sigma_{\mu\nu}d\right)  \left(  \bar{\ell}%
\sigma^{\mu\nu}\ell\right)  $ & CPV ($1^{--}$) & Eq.(\ref{NPT8}) &
Eq.(\ref{EI3})\\
Pseudotensor & $\left(  \bar{s}\sigma_{\mu\nu}d\right)  \left(  \bar{\ell
}\sigma^{\mu\nu}\gamma_{5}\ell\right)  $ & CPC ($1^{+-}$) & Eq.(\ref{NPT12}) &
Eq.(\ref{EI3})\\\hline
\end{tabular}
$%
\caption{Summary of the contributions entering the $K_L\rightarrow\pi^0\ell^+\ell^-$ rate, 
and references to the relevant formulas in the text. The CP-property indicates which of the real or imaginary part of 
the respective Wilson coefficient contributes. Interferences occur whenever the $\bar{\ell}%
\ell$ 
pair is produced in the same state. `Helicity-suppressed/allowed' refers to the last four operators.
\label{Table1}%
}
\end{table}%

In the presence of vector and axial-vector interactions only, the two rates
are bounded in the $\mathcal{B}^{e^{+}e^{-}}$--$\mathcal{B}^{\mu^{+}\mu^{-}}$
plane (see Fig.\ref{Fig5}), i.e., at $1\sigma$,
\begin{equation}
0.1\cdot10^{-11}+0.24\mathcal{B}_{\mathrm{V,A}}^{e^{+}e^{-}}<\mathcal{B}%
_{\mathrm{V,A}}^{\mu^{+}\mu^{-}}<0.6\cdot10^{-11}+0.58\mathcal{B}%
_{\mathrm{V,A}}^{e^{+}e^{-}}\;.
\end{equation}
Any signal outside this region is an indication of New Physics FCNC operators
of different structures (baring the possibility of large $\mu-e$ universality
breaking in the V,A currents). We have identified two possible mechanisms.

The first is from helicity-suppressed (pseudo-)scalar operators, as arising in
the MSSM at large $\tan\beta$. These enhance the muonic mode without affecting
the electronic mode. Such interactions would be revealed by measuring the two
rates above the V,A region. Also, comparing with $K_{L}\rightarrow\mu^{+}%
\mu^{-}$ shows, model-independently, that the $K_{L}\rightarrow\pi^{0}\mu
^{+}\mu^{-}$ is more sensitive (besides being cleaner) to these types of
interactions (see the yellow region in Fig.\ref{Fig5}).

The second possibility is from helicity-allowed (pseudo-)tensor interactions,
which could arise for example from tree-level leptoquark interactions. Because
of the phase-space suppression, it is now the electronic mode which is more
affected. These interactions would manifest themselves in a signal
significantly below the V,A region (see the green region in Fig.\ref{Fig5}).
Even assuming the presence of similar contributions for the $K\rightarrow
\pi\nu\bar{\nu}$ modes, there is at present no severe constraint on these effects.

On the other hand, we found that both helicity-suppressed (pseudo-)tensor
interactions and helicity-allowed (pseudo-)scalar interactions should not lead
to observable effects. For the former, this is so because of significant
phase-space suppression, while the latter are already strongly bounded by the
very rare $K_{L}\rightarrow e^{+}e^{-}$ decay. Also, concerning the
electromagnetic tensor operator, $\left(  \bar{s}\sigma_{\mu\nu}d\right)
F^{\mu\nu}$, measurements of the $K_{L}\rightarrow\pi^{0}\ell^{+}\ell^{-}$
rates cannot disentangle it from vector operator contributions.

For the (integrated) forward-backward asymmetry $A_{FB}^{\mu}$, the first
reliable estimate has been obtained. It is typically of a few tens of
percents, and can give important information, complementary to the total
rates. In the SM, it could fix the sign of the interference between the vector
operator and ICPV contributions. Similarly, beyond the SM, it can be used to
discriminate among various solutions once both $K_{L}\rightarrow\pi^{0}%
\ell^{+}\ell^{-}$ rates are measured. On the other hand, the electronic
asymmetry $A_{FB}^{e}$ is found either completely dominated by its (unknown)
SM value, or too small to be of any use to constrain either New Physics or
$a_{S}$.

We have not included differential rates or differential asymmetries in the
present study (though they can be trivially computed from our analyses). New
physics does affect these observables, but they require a higher experimental
sensitivity, so total rates and integrated asymmetries are more promising.

With the general expressions for both $K_{L}\rightarrow\pi^{0}\ell^{+}\ell
^{-}$ rates and asymmetries computed in Sections 2 and 3, the way is now paved
for more model-dependent analyses. In this context, the $\mathcal{B}%
^{e^{+}e^{-}}$--$\mathcal{B}^{\mu^{+}\mu^{-}}$ plane remains as a particularly
convenient phenomenological tool to display the correlations among operators
specific to a given model.

In conclusion, the $K_{L}\rightarrow\pi^{0}\ell^{+}\ell^{-}$ system, together
with the neutrino modes $K\rightarrow\pi\nu\bar{\nu}$, has a considerable
potential for unveiling/constraining the nature of possible New Physics in
$\Delta S=1$ FCNC, therefore playing an important role in the quest for a
better understanding of the quark flavor sector.

\subsection*{Acknowledgements}

We are pleased to thank Gino Isidori for stimulating discussions and for
reading the manuscript, and Paride Paradisi for useful comments. This work is
partially supported by IHP-RTN, EC contract No.\ HPRN-CT-2002-00311
(EURIDICE). The work of C.S. is also supported by the Schweizerischer
Nationalfonds; S.T.~acknowledges the support of the DFG grant No.~NI 1105/1-1.

\appendix                                                

\section{\textbf{Constraints on (pseudo-)scalar operators from} $K_{L}%
\rightarrow\mu^{+}\mu^{-}$}

We consider the following effective Hamiltonian:%
\begin{equation}
\mathcal{H}_{eff}=\frac{G_{F}\alpha}{\sqrt{2}}\lambda_{t}y_{7A}^{\prime
}\left(  \bar{s}\gamma_{\mu}\gamma_{5}d\right)  \left(  \bar{\ell}\gamma^{\mu
}\gamma_{5}\ell\right)  +\frac{G_{F}^{2}M_{W}^{2}}{\pi^{2}}\frac{m_{s}m_{\ell
}}{M_{W}^{2}}\left[  y_{P}^{\prime}\left(  \bar{s}\gamma_{5}d\right)  \left(
\bar{\ell}\gamma_{5}\ell\right)  +y_{S}^{\prime}\left(  \bar{s}\gamma
_{5}d\right)  \left(  \bar{\ell}\ell\right)  \right]  +h.c. \label{Kmm1}%
\end{equation}
In the SM, $y_{7A}^{\prime}=y_{7A}$ and $y_{P,S}^{\prime}$ are negligible. Our
goal is to get an order of magnitude estimate of the bounds set on
$y_{P,S}^{\prime}$ by the measured $K_{L}\rightarrow\mu^{+}\mu^{-}$ rate.

Using the matrix element parametrizations%
\begin{equation}
\langle0|\bar{s}\gamma_{\mu}\gamma_{5}d|K^{0}\left(  P\right)  \rangle
=i\sqrt{2}F_{K}P_{\mu},\;\;\langle0|\bar{s}\gamma_{5}d|K^{0}\rangle=-i\sqrt
{2}F_{K}\frac{M_{K}^{2}}{m_{s}+m_{d}}\;, \label{Kmm3}%
\end{equation}
in the same conventions as Eq.(\ref{Eq4}), the decay amplitudes $\langle
\ell^{+}\ell^{-}|-\mathcal{H}_{eff}|K_{L,S}\rangle$ and the total rates can be
written as%
\begin{align}
\mathcal{M}\left(  K_{L,S}\rightarrow\ell^{+}\ell^{-}\right)   &
=2i\frac{G_{F}^{2}M_{W}^{2}F_{K}M_{K}}{\pi^{2}}r_{\ell}\left\{  \bar{u}_{p}%
\left(  A_{L,S}^{\ell}+B_{L,S}^{\ell}\gamma_{5}\right)  v_{p^{\prime}%
}\right\}  \;,\label{Kmm4}\\
\Gamma\left(  K_{L,S}\rightarrow\ell^{+}\ell^{-}\right)   &  =\frac{G_{F}%
^{4}M_{W}^{4}F_{K}^{2}M_{K}^{3}}{2\pi^{5}}r_{\ell}^{2}\sqrt{1-4r_{\ell}^{2}%
}\left(  \left(  1-4r_{\ell}^{2}\right)  \left|  A_{L,S}^{\ell}\right|
^{2}+\left|  B_{L,S}^{\ell}\right|  ^{2}\right)  \;, \label{Kmm5}%
\end{align}
with $A_{S},B_{L}$ ($A_{L},B_{S}$) the CP-conserving (CP-violating) pieces
given by%
\begin{align}
A_{L}^{\ell}  &  =\frac{M_{K}^{2}}{M_{W}^{2}}i\operatorname{Im}y_{S}^{\prime
},\;B_{L}^{\ell}=\frac{M_{K}^{2}}{M_{W}^{2}}\operatorname{Re}y_{P}^{\prime
}-\left(  2\pi\sin^{2}\theta_{W}\right)  \left(  \operatorname{Re}\left(
\lambda_{t}y_{7A}^{\prime}\right)  +\operatorname{Re}\lambda_{c}y_{c}\right)
+A_{L\gamma\gamma}^{\ell}\;,\label{Kmm6}\\
A_{S}^{\ell}  &  =\frac{M_{K}^{2}}{M_{W}^{2}}\operatorname{Re}y_{S}^{\prime
}+A_{S\gamma\gamma}^{\ell},\;B_{S}^{\ell}=\frac{M_{K}^{2}}{M_{W}^{2}%
}i\operatorname{Im}y_{P}^{\prime}-\left(  2\pi\sin^{2}\theta_{W}\right)
i\operatorname{Im}\left(  \lambda_{t}y_{7A}^{\prime}\right)  \;. \label{Kmm7}%
\end{align}
The $c$-quark contribution is negligible for $K_{S}\rightarrow\mu^{+}\mu^{-}$,
while for $K_{L}\rightarrow\mu^{+}\mu^{-}$, it has been computed recently to
NNLO giving $y_{c}=\left(  -2.0\pm0.3\right)  \cdot10^{-4}$%
\cite{GorbahnHaisch}. Indirect CPV contributions are understood,
$\mathcal{M}\left(  K_{L,S}\rightarrow\ell^{+}\ell^{-}\right)  _{\mathrm{ICPV}%
}=\varepsilon\mathcal{M}\left(  K_{1,2}\rightarrow\ell^{+}\ell^{-}\right)  $.

The two-photon term $A_{S\gamma\gamma}^{\ell}$ is given in \cite{EckerPich91}
in terms of the two-loop form-factor of Eq.(\ref{Eq9a}):%
\begin{equation}
A_{S\gamma\gamma}^{\mu}=-\frac{\alpha_{em}^{2}G_{8}F_{\pi}}{2G_{F}^{2}%
M_{W}^{2}F_{K}}\left(  1-r_{\pi}^{2}\right)  \mathcal{I}\left(  r_{\mu}%
^{2},r_{\pi}^{2}\right)  =2.10\cdot10^{-4}\left(  -2.821+i1.216\right)  \;.
\label{Kmm8}%
\end{equation}
For $K_{L}$, the situation is less clear as the dispersive part of
$A_{L\gamma\gamma}^{\ell}$ is notoriously difficult to evaluate. Anyway,
following the analysis of Ref.\cite{IsidoriU03}, one can get the conservative
estimate%
\begin{equation}
A_{L\gamma\gamma}^{\mu}=\sqrt{\frac{4\pi^{3}\alpha_{em}^{2}\Gamma\left(
K_{L}\rightarrow\gamma\gamma\right)  }{G_{F}^{4}F_{K}^{2}M_{W}^{4}M_{K}^{3}}%
}\left(  \chi_{disp}+i\chi_{abs}\right)  =\pm1.98\cdot10^{-4}\left(  \left(
0.71\pm0.15\pm1.0\right)  -i5.21\right)  \;. \label{Kmm9}%
\end{equation}
The sign of this contribution depends on the sign of the $K_{L}\rightarrow
\gamma\gamma$ amplitude\cite{IsidoriU03}, itself depending on the sign of an
unknown low-energy constant (see \cite{GerardST}).

To get an order-of-magnitude estimate of the coefficients, we allow for both
signs in the $K_{L}\rightarrow\mu^{+}\mu^{-}$ branching ratio%
\begin{equation}
\mathcal{B}\left(  K_{L}\rightarrow\mu^{+}\mu^{-}\right)  =\left(  6.7+\left(
0.08\operatorname{Im}y_{S}^{\prime}\right)  ^{2}+\left(  0.10\operatorname{Re}%
y_{P}^{\prime}+1.1y_{7A}^{\prime}-0.2\pm0.4_{-0.5}^{+0.5}\right)  ^{2}\right)
\cdot10^{-9}\;, \label{Kmm9b}%
\end{equation}
and reflect only the error on $\chi_{disp}$. Then, imposing the rate to be
within $3\sigma$ of the experimental value $\mathcal{B}\left(  K_{L}%
\rightarrow\mu^{+}\mu^{-}\right)  ^{\exp}=\left(  6.87\pm0.12\right)
\cdot10^{-9}$\cite{PDG} corresponds to $\left|  \operatorname{Im}y_{S}%
^{\prime}\right|  \lesssim8$ and $\left|  \operatorname{Re}y_{P}^{\prime
}\right|  \lesssim35$ for the SM value $y_{7A}^{\prime}=-0.68$. Relaxing this
latter constraint to $\left|  y_{7A}^{\prime}\right|  \lesssim3$ corresponds
to $\left|  \operatorname{Re}y_{P}^{\prime}\right|  \lesssim70$, much larger
since the two interfere in the rate.

For $K_{S}\rightarrow\mu^{+}\mu^{-}$, the experimental bound $\mathcal{B}%
\left(  K_{S}\rightarrow\mu^{+}\mu^{-}\right)  ^{\exp}<3.2\cdot10^{-7}$ is
still very far from the predicted rate of about $4\cdot10^{-12}$ in the SM.
Taking $\left|  \operatorname{Re}y_{S}^{\prime}\right|  ,\left|
\operatorname{Im}y_{P}^{\prime}\right|  \lesssim50$ cannot enhance the
branching ratio beyond $\mathcal{B}\left(  K_{S}\rightarrow\mu^{+}\mu
^{-}\right)  \lesssim10^{-10}$.

Finally, for completeness, the longitudinal muon polarization in
$K_{L}\rightarrow\mu^{+}\mu^{-}$ is expressed as \cite{Herczeg}%
\begin{equation}
P_{L}=\frac{2\sqrt{1-4r_{\mu}^{2}}\operatorname{Im}\left(  B_{L}^{\mu\,\ast
}A_{L}^{\mu}\right)  }{\left(  1-4r_{\mu}^{2}\right)  \left|  A_{L}^{\mu
}\right|  ^{2}+\left|  B_{L}^{\mu}\right|  ^{2}}\;. \label{Kmm10}%
\end{equation}
In the SM, this quantity is entirely driven by the indirect CP-violating
contribution, proportional to $\varepsilon$, and is thus rather small,
$|P_{L}|\sim2\times10^{-3}$\cite{EckerPich91}. Including the scalar and
pseudoscalar currents, we find that for $\left|  \operatorname{Re}%
y_{P}^{\prime}\right|  ,\left|  \operatorname{Im}y_{P}^{\prime}\right|
\lesssim75$, only a 10\% deviation of $P_{L}$ from its SM value can be
generated. On the other hand, $\left|  \operatorname{Re}y_{S}^{\prime}\right|
\lesssim75$ can enhance $\left|  P_{L}\right|  $ up to about $10^{-2}$, while
for $\left|  \operatorname{Im}y_{S}^{\prime}\right|  \lesssim8$, $\left|
P_{L}\right|  $ can be as large as $8\%$. It is indeed this latter parameter
which is the most important since it does not require any $\varepsilon$
factor. This makes $P_{L}$ particularly sensitive to the presence of new
CP-violating sources in the scalar operator (as discussed e.g. in
\cite{ChoudhuryGG00}). Unfortunately, a percent level measurement of $P_{L}$
in the medium term is unlikely, and the $K_{L}\rightarrow\pi^{0}\mu^{+}\mu
^{-}$ total rate is more promising to get a signal or set limits on these New
Physics interactions.

\section{\textbf{Numerical representation of the two-loop form factor}}

In Refs.\cite{EckerPich91,IsidoriSmithUnter}, the two-loop form-factor is
expressed as a complicated three-dimensional integral. For practical purposes,
the following numerical representations can be used instead:%
\begin{align*}
\mathcal{I}\left(  \frac{r_{\ell}^{2}}{z},\frac{r_{\pi}^{2}}{z}\right)   &
=\left\{
\begin{array}
[c]{l}%
\sum_{k=0}^{5}a_{k}^{\ell}\left(  0.33-z\right)  ^{-k/2},\;\;0\leqslant
z<0.315\,,\\
\sum_{k=0}^{5}b_{k}^{\ell}\left(  z-0.30\right)  ^{-k/2},\;\;0.315\leqslant
z\lesssim0.6\;,
\end{array}
\right. \\
\mathcal{I}\left(  \frac{r_{\ell}^{2}}{z},\frac{1}{z}\right)   &  =\sum
_{k=0}^{5}c_{k}^{\ell}\left(  1-z\right)  ^{-k/2},\;0\leqslant z\lesssim0.6\;,
\end{align*}
with%
\[%
\begin{tabular}
[c]{cccc}\hline
$k$ & $a_{k}^{e}$ & $b_{k}^{e}$ & $c_{k}^{e}$\\\hline
$0$ & $10.65-i17.24$ & $54.76-i23.60$ & $-26.069-i3.6914$\\
$1$ & $-6.964+i13.31$ & $-53.89-i37.02$ & $92.05+i0.9337$\\
$2$ & $1.312-i2.521$ & $6.832+i24.74$ & $-122.03+i8.3238$\\
$3$ & $-0.6045+i0.3762$ & $0.942-i5.509$ & $76.611-i8.2021$\\
$4$ & $0.113-i0.0381$ & $-0.265+i0.5508$ & $-23.325+i3.0351$\\
$5$ & $-0.0069+i0.00186$ & $0.0154-i0.0206$ & $2.7774-i0.4036$\\\hline
\end{tabular}
\ \ \
\]%
\[%
\begin{tabular}
[c]{cccc}\hline
$k$ & $a_{k}^{\mu}$ & $b_{k}^{\mu}$ & $c_{k}^{\mu}$\\\hline
$0$ & $-8.5486-i0.9184$ & $-5.0369-i1.4840$ & $-9.6037+i3.7637$\\
$1$ & $9.3942-i0.4070$ & $2.8237+i2.9981$ & $26.408-i14.04$\\
$2$ & $-3.6104+i0.8098$ & $-1.2503-i0.7132$ & $-28.934+i19.15$\\
$3$ & $0.7102-i0.1910$ & $0.3303+i0.1108$ & $16.159-i12.15$\\
$4$ & $-0.0696+i0.0181$ & $-0.0412-i0.0100$ & $-4.5365+i3.715$\\
$5$ & $0.00269-i0.000577$ & $0.00194+i0.000389$ & $0.5087-i0.4428$\\\hline
\end{tabular}
\ \ \
\]
Using this parametrization, one can reproduce the rates and asymmetries
typically with an error of about one percent, which is more than sufficient
given the size of the theoretical uncertainty on the $a_{1}^{P}\left(
z\right)  $ distributions.

\end{document}